# On measuring two-party partisan bias in unbalanced states

John F Nagle[1] and Alec Ramsay[2]

1. Carnegie Mellon University, Pittsburgh, PA 15213.  nagle@cmu.edu
2. Seattle, WA, 98119.  alec@davesredistricting.org

**Abstract**

Assuming that partisan fairness and responsiveness are important aspects of redistricting, it is important to measure them. Many measures of partisan bias are satisfactory for states that are balanced with roughly equal proportions of voters for the two major parties. It has been less clear which metrics measure fairness robustly when the proportion of the vote is unbalanced. We have addressed this by analyzing past election results for four states with Democratic preferences (CA, IL, MA, and MD), three states with Republican preferences (SC, TN, and TX) and comparing those to results for four nearly balanced states (CO, NC, OH, and PA). We used many past statewide elections in each state to build statistically precise seats for votes and rank for votes graphs to which many measures of partisan bias were applied. In addition to providing values of responsiveness, we find that five of the measures of bias provide mutually consistent values in all states, thereby providing a core of usable measures for unbalanced states. Although all five measures focus on different aspects of partisan bias, normalization of the values across the eleven states provides a suitable way to compare them, and we propose that their average provides a superior measure which we call composite bias. Regarding other measures, we find that the most seemingly plausible symmetry measure fails for unbalanced states. We also consider deviations from the proportionality ideal, but using it is difficult because the political geography of a state can entangle responsiveness with total partisan bias. We do not attempt to separate intentional partisan bias from the implicit bias that results from the interaction of the map drawing rules of a state and its political geography, on the grounds that redistricting should attempt to minimize total partisan bias whatever its provenance.

# 1.  Introduction

It is well recognized, not only by political scientists and politicians, that redistricting Congress and state legislatures is important. Indeed, ordinary citizens have engaged in drawing maps using free public software.[1] Not surprisingly, there are many criteria that can be considered in judging maps and different people place different weights on the different criteria.  This paper focuses on two of those criteria that we think are especially important, namely, fairness and responsiveness.[2]

Our measures of bias are for total partisan bias, of which overtly intended gerrymandering is just one part.  There is also underlying bias due to political geography which is sometimes called unintentional gerrymandering (Chen and Rodden 2013). That bias is being quantified using ensembles of maps drawn by computers that use the criteria that a state prescribes in its rules.[3] That provides a benchmark that is then subtracted from the bias of the adopted plan to estimate intended gerrymandering bias.  While this is what is traditionally required for court cases (Grofman 2019, McDonald, et al. 2018), we emphasize that unintentional bias is no longer unintentional if it can be reliably demonstrated that it occurs when following the state's rules or other informal criteria.  This bias then becomes systemic bias due to the state's map drawing criteria interacting with its political geography.  If there is such systemic bias, there may still be plans that are outliers in the ensemble that are nonetheless fair, and we believe a redistricting commission should adopt such a plan rather than an average plan.  However, if no relatively fair plan is possible within the state's rules, then we believe the state should change its election laws to make fair plans possible. Even if a state's rules cannot be changed in time, a redistricting commission and concerned map-drawing citizens could still try to minimize total bias and achieve responsiveness.[4] As has been reviewed by Stephanopoulos 2013, numerous states have had language promoting fairness and responsiveness, but these efforts have been hampered by uncertainties in defining and measuring these concepts, and that is what this paper is about.

---

[1] Some examples are 'Dave's Redistricting App' (Bradlee 2020), 'districtr' (MetricGeometryandGerrymanderingGroup 2018) and 'JudgeIt' (Gelman, et al. 2012).  In Pennsylvania, the Committee of 70 has provided Draw the Lines software (Thornburgh 2018) for people to draw maps to compete for substantial prizes; it unfortunately only uses registration at the overly coarse census tract level instead of election results at the precinct level.

[2] Responsiveness is often called competitiveness, but while closely related at the statewide level we prefer responsiveness to voters over competitiveness for parties.

[3] Of course, implementation of the rules in the computer code generally requires weighting of the different criteria, population equality, compactness, splitting political subunits, communities of interest and minority protection. Remarks made by Katz, et al. 2020, 176 and McGann, et al. 2016, 110 regarding the use of computer drawn map ensembles to evaluate bias are especially pertinent.

This poses the question: Is it possible to test whether a proposed redistricting plan for 2021 will be fair and responsive to the voters?  Our approach to this question is to analyze the enacted redistricting plans for 2011 using many statewide election results applied to those maps.[5]  This paper presents evidence that the use of past election results can indeed produce quite precise information that can be used for evaluating a plan before it is implemented.[6]  We suggest that this exercise can be relevant for the next round of redistricting after the 2020 census, at least in those states that have redistricting commissions not intent on political advantage.

Even in states controlled by one party, the findings in this paper may help challenge unfair and unresponsive plans in court. However, we again emphasize that our work is not focused on challenging plans in courts that, in lieu of precise laws regarding fairness and responsiveness, typically require evidence of partisan intent to overturn a map.[7]  Political geography, such as high density of Democrats in cities, can also create unfair and unresponsive maps when conventional redistricting criteria such as compactness and not splitting political subdivisions are adhered to (Rodden 2019, Nagle 2019).  Our goal in this paper is to elucidate measures of total bias and unresponsiveness, whatever their provenance.[8]  Map drawers can then choose how to balance these metrics with other criteria[9] and a state legislature can be better informed about whether its state election law should be modified.

---

[4] This echoes the discussion of McGann, et al. 2016, 222 that "Electoral fairness is not something that occurs 'naturally'; it has to be actively pursued" and Altman and McDonald 2018,108 that a solution may be to explicitly incorporate political goals into the redistricting criteria.

[5] Statewide elections are the same for all precincts whereas results for Congress or state legislatures are subject to incumbency differences as well as uncontested elections.

[6] Although this conclusion will not come as a surprise to many political scientists, *e.g.* (McGann, et al. 2016, McDonald, et al. 2018, Gelman, et al. 2012, Grofman and King 2007,Stephanopoulos and McGhee 2018,Wang, et al. 2018), the methods and evidence presented here may still be of interest, especially for the express focus on unbalanced states.

[7] See Best, et al. 2020 for insightful concerns regarding reliance on a partisan intent standard.

[8] Following the authors in the previous footnote 6. In particular, "the absence of intentional unfairness is not the same as fairness" Katz, et al. 2020,170.

[9] Criteria balancing has recently been discussed by Altman and McDonald 2018 .

For states that are evenly balanced between two dominant parties, a fair plan is clearly one that is likely to result in half the seats for half the votes.[10] It has been rather more challenging to decide what a fair plan is for a state in which one party routinely obtains considerably more than half the votes (McGhee 2014, Wang 2016, Nagle 2017). It has appeared reasonable to assert that symmetry is required in the sense that if party A wins fraction $S_A$ of the seats with fraction V of the vote, then party B should win the same fraction $S_B = S_A$ of the seats if it received the same vote fraction (*e.g.,* McGann, et al. 2016, 56). A difference in seats $\Delta S = S_A - S_B$ between these two seat fractions would then appear to be a reasonable measure of bias at vote V, with $\Delta S = 0$ being no bias. This has been recently named the β measure of bias (Katz, et al. 2020, 166). It obtains the value $\Delta S(<V>)$, where the $<V>$ means that we calculate $\Delta S$ from the most likely statewide vote $<V>$ determined by the average over many statewide contests and from its counterfactual counterpart when the vote fraction is reversed to $1 - <V>$.[11]

In order to calculate $\Delta S(<V>)$, one must obtain a seats-votes curve S(V) which estimates the fraction of seats S for any statewide vote fraction V (Katz, et al. 2020, 165). We have calculated S(V) curves for the Congressional plans of 11 states. By using many past election results, our S(V) curves have quite small estimated uncertainties. Details of our methodology are presented in Appendix A.

The seats-votes curves are quite interesting, and they are appropriate for evaluating the responsiveness of a plan. However, Section 3.2 of this paper shows that the β measure of bias is highly misleading for states with a dominant party. This came as a shock to us, as symmetry and this way of evaluating it would appear to be so fundamental to what should be considered fair. We will show how and why the β measure fails by examining the vote shares $v_j$ for congressional districts j in the form of a different kind of graph that we call rank-vote r(v) graphs. The r(v) graphs are also quite precise because we again use many election results. They are presented in Sections 2-4.

---

[10] Technically, we estimate the voter preferences of the districts in a plan. These should be balanced such that half the seats would be won if equally attractive candidates are nominated by the parties when the overall statewide vote is split evenly. Of course, these conditions are unlikely to be met in any real district which is why actual elections are important, so our methods are therefore not designed to predict outcomes for individual districts. Nevertheless, for states with many districts, such influences tend to average out, so it becomes possible to predict overall outcomes more reliably. Even so, that is not the goal. Rather, as mentioned by Cervas and Grofman 2020,7, it is to test the fairness and responsiveness of a plan *ab initio*, before specific candidates are chosen and before people and parties decide how to allocate resources to the various contests. It is also certainly not the goal to predict the vote fraction in a future election, only to estimate what the overall preference would be given the overall vote fraction V.

[11] Katz, et al. 2020 define β(V) for all V and prefer to evaluate at $V = <V>$, the most probable statewide vote or a range including $<V>$. For succinctness, we will call this the β measure.

The r(v) graphs together with the S(V) graphs enable us better to portray and understand bias. Nevertheless, as has been well known for a long time, there is still ambiguity in the single member district system as to what is fair in states that have a dominant political party. In such unbalanced states, bias and responsiveness become intertwined (McGann, et al. 2016,67). For example, if the two-party political balance in a state is 60/40, then a plan that has each district with a 60/40 preference would likely elect all its representatives from the same party. That would be fair under the aforementioned β symmetry principle.[12] It may also be the only possibility if the political geography of the state is completely homogeneous.[13] However, if the state's political geography were completely heterogeneous, one could draw a map that guarantees a 60/40 split in the seats,[14] but one could also combine the precincts to give all seats to the dominant party or to any split between these extremes.

Whatever one deems fair in the preceding examples, they illustrate that the political geography of a state could clearly be important for assessing bias and responsiveness in unbalanced states (Rodden 2019). We examine this further in Appendix C where we recall the ideal of proportionality in Section C.1. This leads us in Section C.2 to a method to measure bias in unbalanced states that takes political geography into account, and a qualitative analysis of states is given in Section C.3. However, that approach is difficult to apply to any single plan, so the main line of analysis in this paper proceeds differently in Section 5. Global symmetry is defined in subsection 5.1 Subsection 5.2 defines a completely new measure (γ) of bias, followed by subsections elaborating on the declination (δ) measure (Warrington 2017) and the lopsided outcomes (LO) measure (Wang 2016,1263) and concluding in Section 5.5 by considering maximal responsiveness as the primary goal. Comparisons of these and more standard measures of bias are given in Section 6 which crucially demonstrates that a core subset consisting of five of these measures of bias is substantially consistent across the 11 states, including unbalanced states, and we propose that suitably averaging their values into what we call composite bias Ω provides a superior estimate of bias. Subsection 6.3 also tests measures of bias for durability as the statewide vote share changes.

---

[12] One would have 40/60 preferences in each district if 1/3 of the dominant party's voters switch and then the other party would win all the seats and therefore β = 0.
[13] If every precinct has the same 60/40 balance, all districts must have the same 60/40 balance.
[14] This is most easily seen by defining complete heterogeneity as 60% of the precincts having only dominant party voters and creating districts only from like-minded precincts.

A general discussion ensues in Section 7. Our main result that several measures of bias agree for unbalanced states as well as for balanced states leads us to conclude that total bias can be reliably measured in all states even though the political geography varies by state.

## 2. Analysis of four politically balanced states

This section also describes some of our terminology.

### 2.1 Example of Colorado

We begin with the state of Colorado (CO). The average statewide vote <V> for 12 elections in the time period 2004-2012 was 50.6% Democratic, making this a politically balanced state in that time frame. Figure 1 shows the average two-party Republican vote fraction $v_j$ for each congressional district (CDj, j=1-7) along with the standard error of the mean over 12 elections.[15] The districts are rank ordered according to their placement j in a list that is ordered by the vote. We call this kind of graph the rank r(v) graph because it plots the district rank versus the votes $v_j$ in the districts.[16] According to the r(v) graph in Fig. 1, CD1 and CD2 were strongly Democratic and CD4 and CD5 were strongly Republican in the 2011 CO map, whereas CD3, CD6, and CD7 were more competitive.[17]

---

[15] Following McDonald, et al. 2018, we used data from Wolf 2014. Lists of the chosen elections for all states are given in Appendix A, along with a brief description of Wolf's data. Along with others (Backstrom, et al. 1990, Gronke and Wilson 1999, McDonald 2014, Cervas and Grofman 2020, Powell, et al. 2020, Abramowitz, et al. 2006) for the purpose of determining political preference, we prefer statewide elections data over actual congressional elections which do not uniformly evaluate precincts in different congressional districts, partly because of incumbency and uncontested elections. Also, there are more statewide races for more precise statistics.

[16] It is convenient to express the rank axis r as (j - ½) divided by the number of districts in the state, where j is the rank in the ordered list, so the rank axis uniformly spans the range 0 to 1 for all states.

[17] Indeed, districts 1, 2, 4, and 5 remain with the same party when any of the statewide elections are applied to them, whereas districts 3, 6, and 7 switched parties for different elections. Note that the uncertainty bars on the r(v) data in Fig. 1 are standard errors of the mean which indicate the precision of the partisan preferences of the districts. The standard deviations are the square root of 12 (number of elections) greater; while the standard deviation would better show the switching of parties in CD3 and CD7 from different statewide elections, the standard error of the mean better indicates the precision of the district preferences.

An r(v) graph displays the partisan preferences of each district in a state. It is also often used to count the partisan number of seats in the following too simple way. In Fig. 1, one might say that preferences[18] in CO are for three Democratic seats, in districts 1, 2 and 7, and four Republican seats, in districts 3-6. However, that 'all-or-nothing' assignment of the integers 1 or 0 to each district doesn't take into account uncertainties in competitive districts like CD6. Instead of 'integer assignment', we assign probabilities to districts thereby obtaining fractional seats.[19] Furthermore, Fig. 1 is for only one overall statewide vote share $<V_R> = 49.4\%$ Republican, the complement of the Democratic two-party vote share. What is also needed is a different kind of figure, namely, the seats-votes S(V) graph, which estimates the fraction of seats S for any two-party statewide vote fraction V. We use the proportional shift method to adjust each district vote $v_i$ as the statewide vote shifts.[20] Our S(V) graph for CO is shown in Fig. 2.

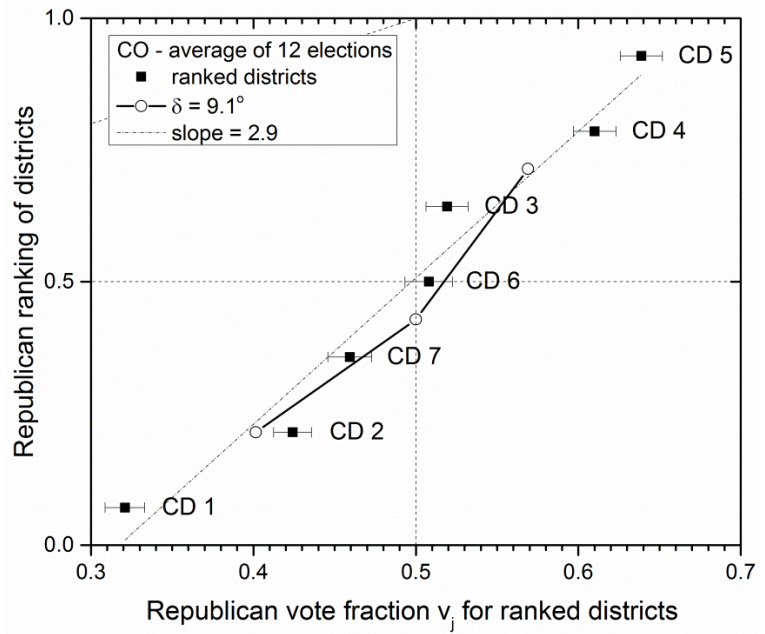

Fig. 1. Colorado r(v) graph for the 2011 plan shows Republican district rank vs. Republican district vote. The slope of the linear fitted line to the districts is 2.9. The horizontal uncertainty bars are standard errors of the mean. The declination is the difference in the angles of the two solid lines *(vide infra)*.

---

[18] Notice that we do not write that CO would be expected to elect three Democrats. By using statewide election results, we deliberately do not take incumbency in a district into account because the goal is to evaluate the partisan preferences of the 2011 map independently of political contingencies.

[19] For example, a district that has a 50/50 partisan preference should count as half a seat for both parties. Details for fractional seat assignment and other technical aspects in this paragraph are given in Appendix A. Fractional seats have been employed in various ways by Gelman and King 1994,532 and Cottrell 2019, and the concept is nicely explained by McGann, et al. 2016, 58-60 where the 5% variation that is employed in our study was suggested.

[20] Although a proportional shift (Nagle 2015, 2019) is conceptually superior to the commonly used uniform shift, the difference between the two is not consequential for the results in this paper as is shown in Appendix B.

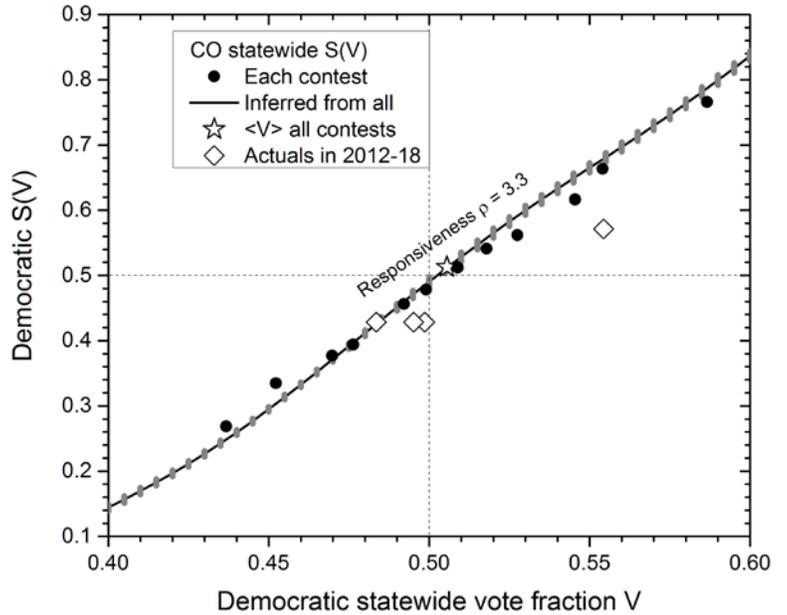

Fig. 2. S(V) graph showing Democratic seat fraction preference vs. Democratic votes fraction curve for the Colorado 2011 enacted plan as a continuous solid curve. Each election ε generated an $S_\varepsilon(V)$ curve (not shown) using proportional shift and fractional seats, and the shown solid S(V) line is the average with standard error of the mean uncertainties as solid vertical bars. The solid circles are the results of each statewide election applied to the plan. The star shows the estimated seat fraction for the average statewide vote <V> for all elections. The diamonds show the fraction of seats won in the actual congressional contests carried out under the 2011 map.

An important result from S(V) curves is the responsiveness of a map. How many seats are expected to change as the statewide vote V changes around V is just the slope dS/dV of the S(V) curve at V. Of course, the slope varies with V if the curve is not linear and then the most relevant measure of responsiveness is $\rho = dS/dV$ at V = <V>. Because the S(V) curve for CO is quite linear, however, its responsiveness is nearly the same for all V. An important finding for the CO 2011 plan is that its responsiveness is quite high, namely $\rho = 3.3$.[21] This is far higher than the responsiveness $\rho = 1$ idealized by proportionality and even for the efficiency gap (EG) (McGhee 2014, Stephanopoulos and McGhee 2018) which idealizes responsiveness of $\rho = 2$. It is closer to the classical cube "law" (Kendall and Stuart 1950) that has $\rho = 3$ at V=0.5.

Each solid circle in Fig. 2 shows an unshifted result from one election. Already, those points could be simply fit to provide a reasonable S(V) curve. Our method is more precise, as is indicated by the standard errors of the mean shown as vertical bars on the S(V) curve.[22] The diamonds in Fig. 2 show actual congressional seats for the 2012-2018 district elections. These agree with the S(V) curve as well as could be expected, given that they must be integers in actual elections, and that they are affected by contingencies such as incumbency.

---

[21] This value of ρ means that a swing of 5% in the statewide vote V would change the estimated seat fraction by 16.5%. For CO with 7 districts, that is a net swing of essentially one congressional seat.

[22] Also, our method does not require choosing a fitting function, such as a bilogit King and Browning 1987. Although a linear fit to the circles in Fig. 2 would work well enough for CO, other states are considerably nonlinear.

## 2.2. North Carolina

We turn next to North Carolina (NC), another nearly balanced state with Democratic average vote $<V> = 51.5\%$ in the statewide elections in our data set. Fig. 3 shows its r(v) graph. It is very far from linear as shown by the dash-dot line, which is the best linear fit, so the corresponding slope of 1.85 is meaningless as a measure of responsiveness. Fig. 3 emphasizes the well-known fact that the 2011 map for NC has three heavily packed Democratic districts and ten safe Republican districts that are not packed.

An interesting measure of bias is the declination δ, which Warrington (2017) has defined as the difference in the angles of the two straight lines joining the open circles in Figs. 1 and 3.[23] Packing Democrats into a few districts moves the middle open circle in Fig. 3 to smaller ranking of the districts won by Democrats which increases the differences in the angles of the lines connecting to the average party votes, so a positive value of δ correlates with an advantage for Republicans. This measure gives a large positive δ for NC, in contrast to a much smaller value for CO.

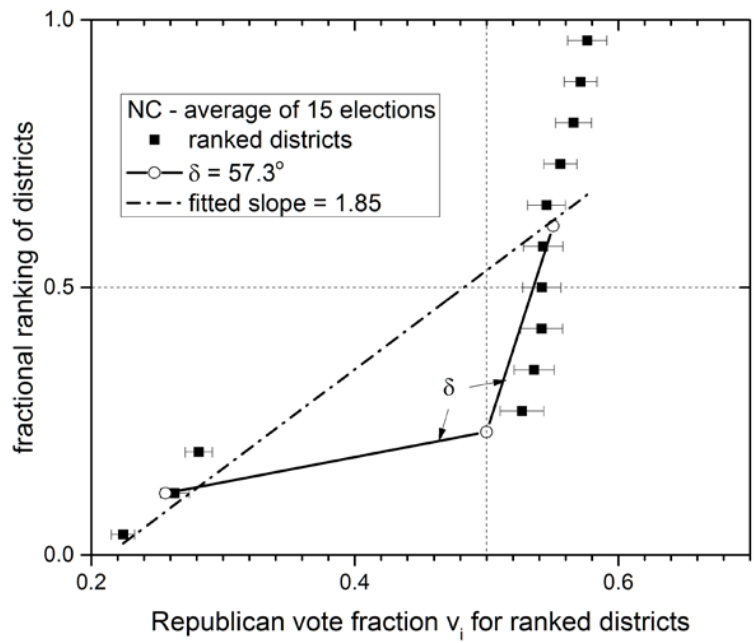

Fig. 3. Republican district rank vs. Republican district vote r(v) graph for North Carolina 2011 plan. The difference in the angles of the two solid lines is the declination δ which equals 57.3° using all-or-nothing seats.

---

[23] Each outer circle locates the average vote for each party's seats and its average district rank. The middle open circle in Figs 1 and 3 locates the rank that divides Democratic won seats from Republican seats. More details regarding this measure are provided in section 5.3.

Fig. 4 shows the S(V) curve for NC.[24] As it is derived from the data in Fig. 3, it too is highly non-linear. Again, the individual statewide election results shown by solid circles agree well with the S(V) curve. So do the actual congressional election results for 2012-2018, but these also show that NC voters tended to vote more conservatively for congress than for the statewide elections; the latter average <V> is shown by the star whereas the range of the actual congressional statewide vote average is centered near 0.48. This is important for evaluating responsiveness. The slope of the curve at <V> is quite high, $\rho = 4.0$ in Fig. 4. However, the slope is close to zero near V = 0.45 which is the low end of the actual range of congressional elections, thereby indicating an unresponsive map favoring Republican incumbents in ten districts.

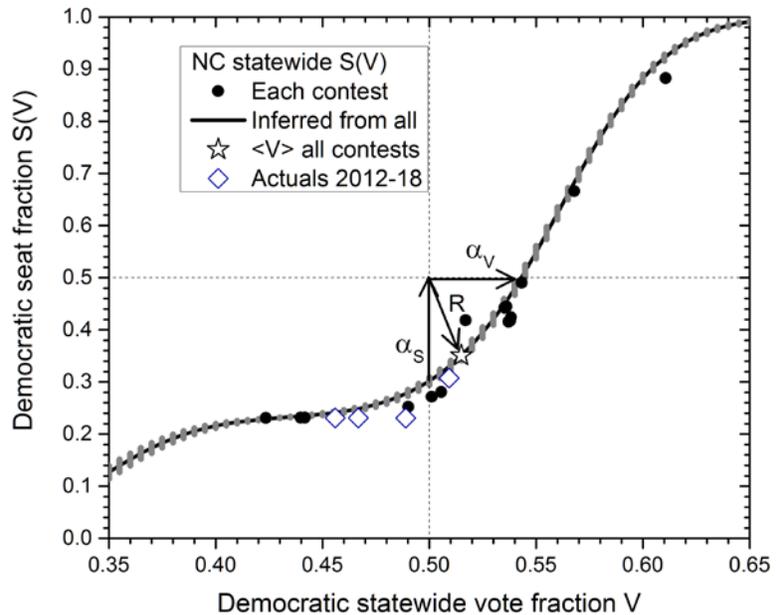

Fig. 4. Seats-Votes S(V) graph for the North Carolina 2011 enacted plan. The solid S(V) curve was obtained using proportional shift and fractional seats. Each contest is the result of each statewide election applied to the map. The star shows the average statewide vote for all elections. Diamonds show actual outcomes. The arrows indicate measures defined in the text.

We will use Fig. 4 to illustrate our definitions of several other quantities of interest. One such is the slope R of the arrow pointing from the center of the graph to the star on the curve.[25] A negative value of R = -10 signifies an anti-majoritarian result, fewer than half the seats for more than half the vote.[26] For unbalanced states R is positive and can be thought of as a winner's bonus. For want of a better name, it may be thought of as an overall responsiveness.

---

[24] The reader may wonder why we have chosen to use Republican rank and votes axes in our r(v) graphs and Democratic seats and votes in our S(V) graphs. The reason is that doing so makes the r(v) and S(V) curves look more alike. For example, the median district with rank 0.5 has R vote greater than 0.5 in Fig. 3 and the Democratic vote V to obtain half the seats is greater than 0.5 by very nearly the same amount. Indeed, using all-or-nothing seats for each district and uniform shift, the r(v) curve with R axes becomes an S(V) curve with D axes.

[25] $R = (S(<V>) - \frac{1}{2})/(<V> - \frac{1}{2})$. It is quite different from responsiveness $\rho$ which is the tangent to the S(V) curve given by the derivative dS/dV at <V>.

An important seats-based measure, much employed in the literature, is the seats bias defined as the difference in the fraction of seats from 0.5 when V = 0.5. In this paper we will designate this measure by $\alpha_S$.[27] Its magnitude is the length of the vertical arrow in Fig. 4, favoring Republicans by 20%. An important votes-based measure will be designated $\alpha_V$; it focuses on the estimated fraction of the vote for 50% of the seats. In Fig. 4, 54.3% of the vote would have to be won by the Democrats in NC to win half the seats. We designate the $\alpha_V$ value to be this vote at S = ½ minus 0.5; it is the length (in %) of the horizontal arrow in Fig. 4.[28]

### 2.3. Other balanced states

The results in the previous section clearly identify the 2011 NC map as unfair to Democrats and unresponsive to voters. We have also drawn r(v) and S(V) graphs for the 2011 maps of two additional fairly evenly balanced states, Pennsylvania and Ohio.[29] The $\alpha_S$, $\alpha_V$, $\delta$ and R values for PA and OH are rather similar to NC in being unfair and anti-majoritarian, and quite unlike the results for CO which appear to be both fair and responsive.

Table 1 collects numbers for the quantities of interest mentioned in this section for balanced states. This table also shows values of quantities that will be defined in subsequent sections. It further exhibits results for several unbalanced states to which we turn in the next two sections. Uncertainties in many of the quantities can be discerned by the reader from the error bars in the S(V) graphs. For example, typical uncertainties in $\alpha_S$ are less than 1%. Uncertainties have not been estimated for $\delta$; the values in Table 1 were calculated using fractional district seats and votes which differ from the values in the legends of our r(v) graphs which used the original definition of Warrington, 2017 that used all-or-nothing seat assignment.

---

[26] The 2018 actual congressional election also was anti-majoritarian as shown by one of the diamonds in Fig. 4.
[27] $\alpha_S$ is also the same as β (0.5).
[28] The goal of the $\alpha_V$ metric, which is based on the S(V) graph, is similar to that of the popular median minus mean metric (mM) McDonald and Best 2015 which is based on the data in the r(v) graph.
[29] These and graphs for other states will be available in SM.

| state | <V> | S(<V>) | R | ρ | ζ | α1 | α2 | δ | GS | γ | Ω | β | LO | PR | EG |
|---|---|---|---|---|---|---|---|---|---|---|---|---|---|---|---|
| MD | 59.3 | 85.7 | 3.7 | 1.0 | 0.8 | -5.2 | -1.0 | -33 | -2.6 | -26 | -2.4 | 1.4 | 4.1 | -26 | -17 |
| CA | 59.2 | 73.6 | 2.6 | 2.1 | 0.3 | -1.9 | -0.6 | -3 | -1.8 | -4 | -0.6 | 2.9 | 9.2 | -14 | -5 |
| MA | 60.0 | 96.0 | 4.6 | 1.9 | 0.3 | 6.4 | 1.0 | NA | 2.0 | -27 | 1.2 | -2.9 | NA | -36 | -26 |
| IL | 60.0 | 78.8 | 2.9 | 3.1 | 0.1 | 5.8 | 1.6 | 6 | 1.9 | 2 | 1.0 | 1.2 | 11.7 | -19 | -9 |
| CO | 50.6 | 51.3 | 2.2 | 3.9 | 0.2 | 1.1 | 0.3 | 0 | 0.8 | 1 | 0.2 | 1.0 | 0.3 | -1 | -0 |
| PA | 52.9 | 43.6 | -2.2 | 4.1 | 0.1 | 16.1 | 4.4 | 24 | 5.5 | 18 | 3.4 | 14.5 | 8.7 | 9 | 12 |
| NC | 51.5 | 35.1 | -10.0 | 4.0 | 0.1 | 19.8 | 4.3 | 37 | 6.2 | 21 | 4.0 | 19.0 | 11.1 | 16 | 18 |
| OH | 51.3 | 41.4 | -6.0 | 4.5 | 0.1 | 13.8 | 3.1 | 22 | 4.1 | 14 | 2.7 | 13.2 | 6.9 | 10 | 11 |
| SC | 43.0 | 16.1 | 4.9 | 0.9 | 1.0 | 14.1 | 2.3 | 48 | 5.0 | 28 | 3.7 | 0.9 | 4.2 | 27 | 20 |
| TN | 41.6 | 19.2 | 3.6 | 0.8 | 1.1 | 16.2 | 2.9 | 35 | 6.1 | 24 | 3.7 | -2.3 | 0.6 | 22 | 14 |
| TX | 40.4 | 27.8 | 2.3 | 1.1 | 0.7 | 5.0 | 1.2 | 17 | 5.2 | 12 | 1.9 | -7.5 | -3.0 | 13 | 3 |

Table 1. A list of the states and many of the quantities of interest obtained in this paper. Values are expressed as percentages except for the δ angle and the responsiveness measures R, ρ, and ζ. Positive values for measures of bias favor Republicans.

- <V> is the average statewide two-party vote for Democrats.
- S(<V>) is fractional Democratic seats at <V>.
- R is an overall measure of responsiveness or winner's bonus defined in Fig. 4.
- ρ is the slope of the S(V) curve at <V>.
- ζ is a measure of responsiveness defined in section 5.5.
- α1 = $α_S$ is half the difference in party seats at V=0.5.
- α2 = $α_V$ is the excess vote required for half the seats.
- δ is the value of the declination angle calculated using fractional seats and votes.
- GS is a global symmetry measure described in section 5.1.
- γ is the fair difference in party seats at <V> using ρ and defined in section 3.1.
- Ω is the average of the five previous measures, each normalized to the $α_V$ scale.
- β gives the counterfactual symmetry difference in seats, defined in the introduction.
- LO shows values for the lopsided outcomes measure described in section 5.4.
- PR is the deviation from proportionality S(<V>) - <V>.
- EG is the efficiency gap (McGhee 2014) S(<V>) - 2<V> + ½ .
- mM (not shown in table) is the median minus mean (McDonald and Best 2015).

## 3. Republican majority states

This section examines politically unbalanced states with a substantial Republican majority.

### 3.1 Tennessee and the 'Wall'

Tennessee (TN) has an average D vote $<V> = 0.416$. Its $r(v)$ graph in Fig. 5 shows a highly Democratically packed CD9, while CD5 leans Democratic and there are 7 safe Republican districts that figuratively form a 'wall'. The graph is quite non-linear and this and the wall are reflected in the large positive value of the declination angle $\delta$. The corresponding S(V) curve in Fig. 6 agrees well with the statewide races (solid circles) and it agrees as well as can be expected with the actual congressional results given that those are constrained to be integers and that the incumbency effect may have helped retain the second Democratic seat. The values of the seats-based measure $\alpha_S$ and the votes-based measure $\alpha_V$ given in Table 1 are similar to those of the balanced states NC, OH, and PA. This comparison suggests that TN is also biased against Democrats. Unlike those states, R is positive, and there were no anti-majoritarian results (solid circles in Fig. 6) where more than half the votes yielded less than half the seats. The large value R = 3.6 gives a substantial winner's bonus of 2 seats compared to R = 1 proportionality. The much smaller value of the responsiveness $\rho$ compared to all the other states in Table 1 is consistent with the drawing of mostly safe seats for both parties as would be done in a bipartisan gerrymander.

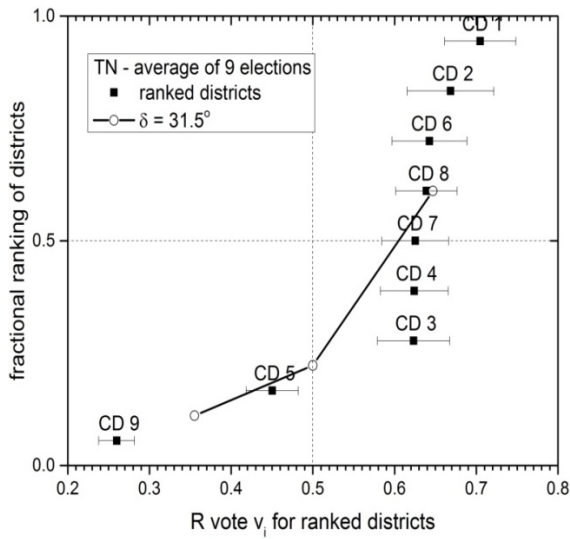
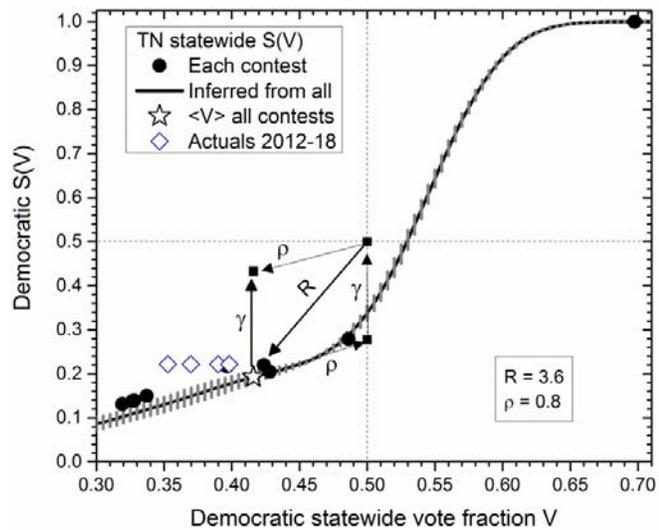

Fig. 5. Rank-vote graph for Tennessee.   Fig. 6. Seats-votes graph for Tennessee.

Figure 6 introduces a new measure of bias which we will identify as γ. It uses the responsiveness ρ evaluated at <V> to calculate a seat fraction $S_ρ$. In Fig. 6, an arrow emanates from the fair point at S=0.5 for V=0.5 with slope ρ. The end of the arrow at V = <V> locates $S_ρ$ which is shown as a solid square in Fig. 6. The seat difference between $S_ρ$ and S(<V>) is defined as γ. Alternatively, Fig. 6 shows an arrow with slope ρ emanating from the S(<V>) point to a projected S(0.5) value, and then γ is the difference compared to the fair S(0.5) = 0.5 value. Because γ uses both responsiveness and seats, it could be described as a 'responsiveness & seats' measure. A full discussion of the merits and characteristics of this measure will be deferred to Section 5.2. Here we will only note that it is computed entirely from data at the statewide vote <V> and so it does not employ counterfactual shifting.

### 3.2. Flaw in the β measure

We now address the use of symmetry to estimate bias according to the measure defined as β in the introduction. The S(V) curve in Fig. 6 gives S(<V>) = 0.192 for Democrats with <V> = 0.416. The counterfactual symmetrically opposite vote is 1 - <V> = 0.584, and at that value on the S(V) curve the fraction of D seats is 0.857, so the counterfactual fraction of R seats is 0.143. As this is less than the factual D seat S(<V>) = 0.192, the symmetric measure of bias β posits that TN is biased in favor of Democrats, contrary to all the other measures. As this differs from expectations, one might question the accuracy of our S(V) methodology.[30] However, the small uncertainties we obtain for our S(V) curves have led us to reconsider the β measure and to conclude that it gets confounded in unbalanced states.

The key to understanding the flaw in the β measure comes from the r(v) graph. If the vote were to shift strongly Democratic, the entire wall of safe Republican districts would fall. This is in contrast to the sole highly packed Democratic district that anchors S(<V>). A simple toy example illustrates this phenomenon. Consider a state with ten districts, nine of which have party A preference of 0.65 and one which has party A preference of 0.25 so the state has an overall preference $<V_A> = 0.61$ for party A.[31] Party A would then be expected to obtain nine seats with this set of preferences. For simplicity let us assume for the counterfactual that each district undergoes a uniform swing to party B by $\Delta V = 0.22$ which achieves the counterfactual statewide party B preference $V_B = 0.61$. Then there are nine districts with party A preference of 0.43 and one district with preference 0.03. For simplicity, let us use all-or-nothing district assignment which would then give party B ten seats in the counterfactual. The $\beta$ measure would then draw the absurd conclusion that this state is biased in favor of party B because it would have obtained more seats if it had counterfactually received the same statewide vote fraction as party A actually received.[32] This illustrates that by treating the two parties differently – most clearly illustrated by asymmetrical r(v) graphs with a significant angle of declination – map drawers can achieve advantage for their party while the simple $\beta$ measure of symmetry suggests the opposite. For a comparison that is relevant for the $\alpha_S$ measure, a uniform shift of 0.11 to $V = 0.5$ would still give 9 seats to party A using all-or-nothing district seat assignment or 7.6 seats using fractional seats.

---

[30] Of course, our value for β was obtained from a model for shifting the vote. The alternative uniform shift model essentially agrees by giving a similarly small value for β, although of the opposite sign. It may also be noted that TN is unusual compared to other Republican majority states in that there was one election, 2006 Governor, that had a strongly Democratic vote share of 0.70. This outlier election is responsible for giving the visibly large uncertainties for 0.3 < V < 0.45 in Fig. 6 and also for the rather larger standard errors of the mean for the districts in Fig. 5 compared to other states. Nevertheless, our inclusion of this election demonstrates the robustness of our S(V) curves over a large range of vote share.

We wish to emphasize that this section only criticizes the $\beta$ measure of bias, not the fundamental concept that fairness requires symmetry, as will become apparent in section 5.1.

### 3.3. Texas and South Carolina

We have also analyzed two other Republican majority states, Texas (TX) and South Carolina (SC). Results shown in Table 1 for the $\alpha_S$ and $\alpha_V$ measures for SC are similar to those for TN and the value of $\delta$ is even higher. The same measures for TX are somewhat smaller, but still biased in favor of Republicans. In contrast, the $\beta$ value is nearly zero for South Carolina and even would assign substantial bias in favor of Democrats in TX, again indicating that the $\beta$ measure is flawed.[33] Included here as Fig. 7 is the r(v) graph for TX. It also exhibits a wall of safe Republican districts in the $0.6 < V < 0.7$ range, although the TX wall is not as steep or high as the TN wall.

---

[31] This assumes that all districts had the same number of voters, *i.e.*, no turnout bias. Effective turnout bias is small in this study as is documented in Appendix A.2.

[32] Using proportional shift and fractional seats slightly alleviates the flaw in the $\beta$ value by giving 9.6 instead of 10 sets to party B in the counterfactual. We also note that other examples reveal that the unfairness that is not diagnosed by the $\beta$ measure comes not from packing minority party districts, but from building a wall of majority party seats that are safe but not packed.

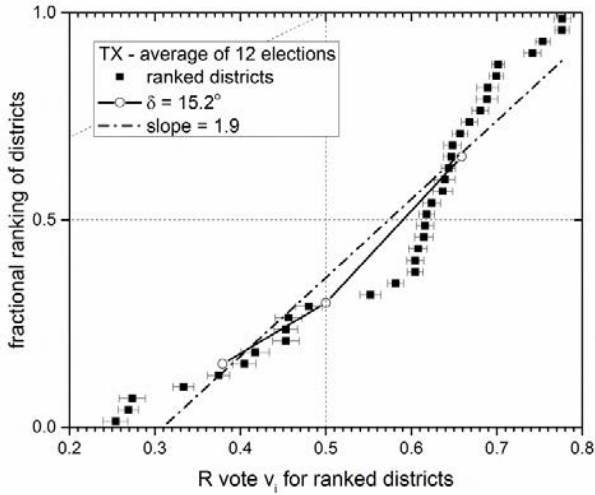 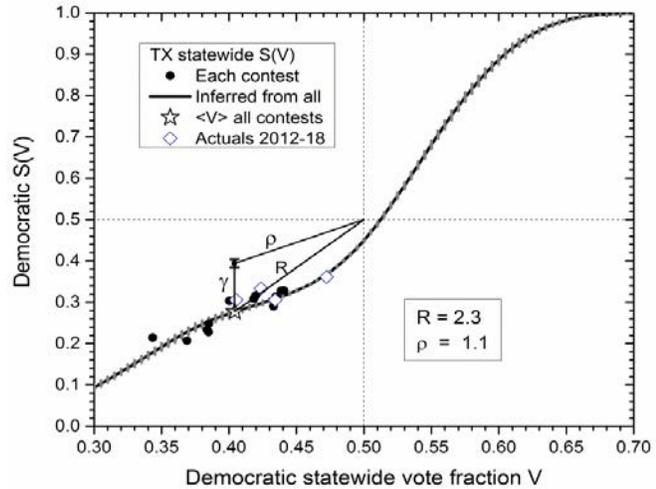

Fig. 7.  r(v) graph for Texas.                     Fig. 8  S(V) graph for Texas.

## 4. Democratic majority states

### 4.1 California and Illinois

It is widely recognized that California (CA) has been a leading state for redistricting reform, so it is especially interesting to examine its congressional plan using our methods. The r(v) graph in Fig. 9 is basically linear with a small value of the declination δ favoring Democrats. The linear fit to the r(v) data gives a slope of 2.1. This is the same as the responsiveness ρ shown in the S(V) graph in Fig. 10.[34] The winner's bonus R = 2.5 for CA is considerably larger than proportionality (R = 1) and somewhat larger than the EG (R = 2), but still considerably smaller than the winner's bonuses in TN and SC. The $\alpha_S$ and $\alpha_V$ measures in CA favor

---

[33] Our data for SC and TX also differ from that of NC in that there was no statewide election won by the Democrats. That means that the counterfactual S(1 - <V>) had to be extrapolated well beyond the vote share range of the available elections, in contrast to the less uncertain interpolation within the range of available vote share that was done for TN. Nevertheless, the general shapes of the extrapolated S(V) curves for SC and TX are quite similar to the one for TN as shown in Fig. 8 for TX.

[34] Our S(V) curve agrees quite well with one drawn by McGann, et al. 2016,78, although our uncertainties are smaller. They also provide values for the $\alpha_S$ measure for all states in their Table 3.A.2. Our value for CA is in excellent agreement and values for other states are also satisfactory considering that we use statewide elections and they used district elections.

Democrats, but less so than they favor Republicans in TX and much less than in TN and SC as shown in Table 1.

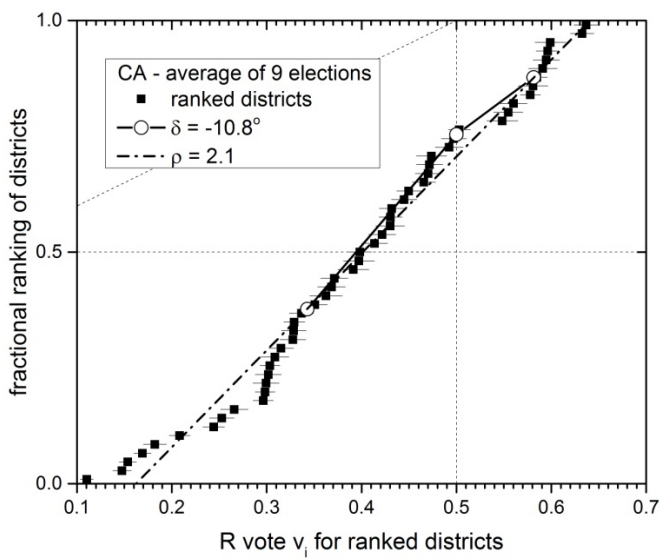
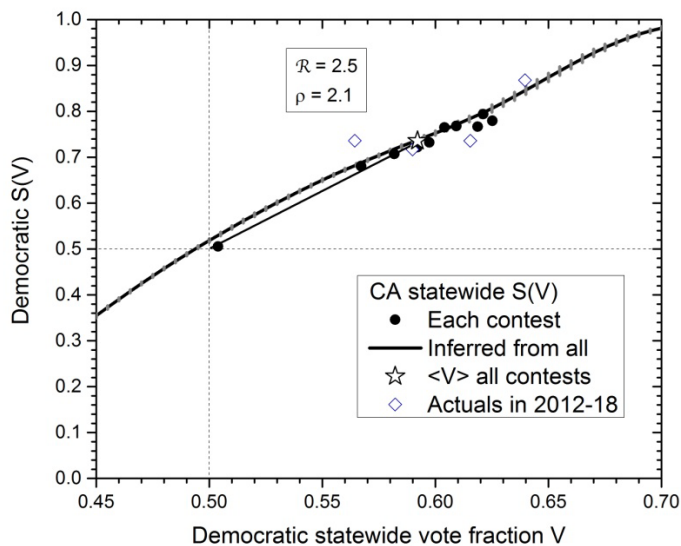

Fig. 9.  r(v) graph for California     Fig. 10.  S(V) graph for California.

We have also analyzed the 2011 Illinois (IL) plan. Table 1 shows that the IL plan is somewhat more responsive than CA; that gives a relatively larger Democratic seat fraction in IL because both states have nearly the same average <V> vote.  Its r(v) graph is not as linear as for CA, but its δ is smaller and even favors Republicans.  There appears to be a soft wall of six Democratic districts centered near 0.43 R vote in the r(v) graph, but this wall appears to be close enough to V = 0.5 such that, along with the extreme packing of Democrats in other Chicago districts, IL has small positive $\alpha_S$ and $\alpha_V$ values favoring Republicans.  (Supplementary material contains the IL graphs.)  We concur with the analysis of McDonald, et al. 2018, 323 who wrote "Despite the outcries of unfairness in some quarters of the press and a court seeing a 'blatant political move' … Democrats' self-help maneuvering was largely a matter of tamping down some of the pro-Republican effects of residential patterns."  Similar comments were written by McGann, et al. 2016, 105.

### 4.2  Maryland and Massachusetts

We now turn to two other Democratic majority states that differ substantially from CA and IL and also from each other.  Maryland (MD) is widely regarded as having been intentionally gerrymandered by Democrats. Its r(v) graph in Fig. 11 shows that the infamous CD1 is a safe Republican district, and CD6 leans substantially Democratic. The complaint is that CD6 could be made more competitive or even leaning Republican if Democratic voters were exchanged with R voters in adjacent CD8. That would shift the middle declination circle down in Fig. 11 which would reduce the very large negative value of δ and it would soften the already soft wall consisting of CD5, CD8, CD2, and CD3.

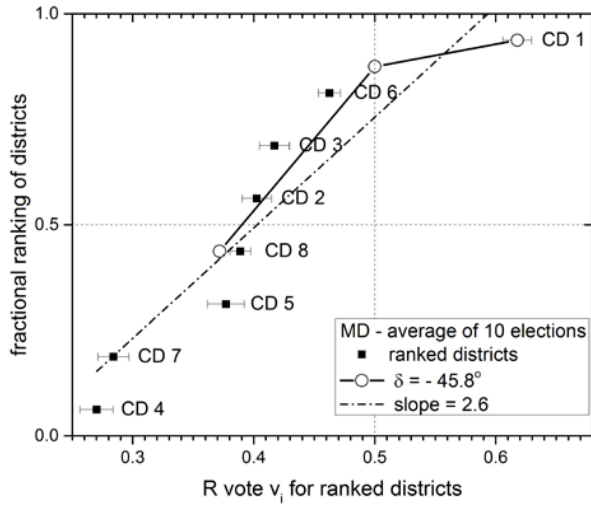 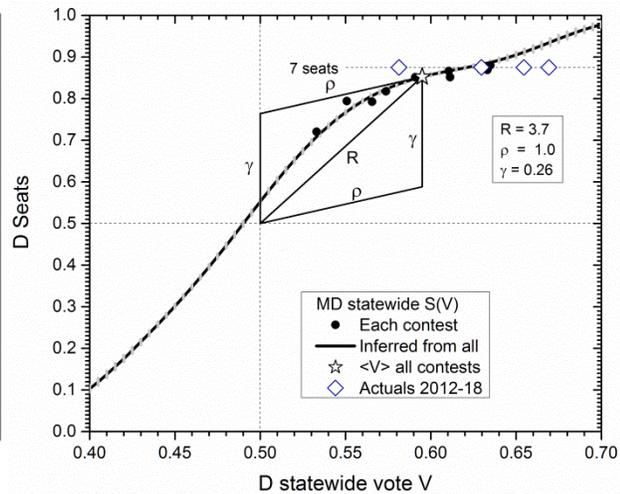

Fig. 11. r(v) graph for Maryland    Fig. 12. S(V) graph for Maryland

The S(V) curve in Fig. 12 shows that the winners bonus R is quite large but the responsiveness ρ is much smaller, consistent with drawing relatively safe districts for both parties. Fig. 12 also shows that the $\alpha_S$ and $\alpha_V$ measures favor Democrats, although Table 1 shows that the magnitudes of these bias values are much smaller than the corresponding values for five states that these measures purport to be biased in favor of Republicans.

Massachusetts (MA) has a similar two-party average vote share in our data compared to MD, but MA has not elected any Republicans to Congress under its 2011 plan whereas MD has always elected one. The r(v) graph for MA in Fig. 13 has a highly packed Democratic district CD7 and a wall of eight Democratic districts with Republican support in the range $0.35 < v_i < 0.45$. A linear fit just to this wall gives a large pseudo responsiveness of 7.3. The S(V) graph in Fig. 14 shows that a large winner's bonus like R = 4.6 results in the dominant party obtaining nearly all the seats when <V> differs from 50% by as little as 10% of the vote. There were ample elections near V = 0.5 to give credibility to the $\alpha_S$ and $\alpha_V$ measures which actually indicate bias in favor of Republicans in the 2011 MA plan.[35] On the other hand, the $\alpha_S$ and $\alpha_V$ measures indicate bias for the Democrats in the MD plan, but that plan provides for one Republican seat whereas the MA plan essentially guarantees none.

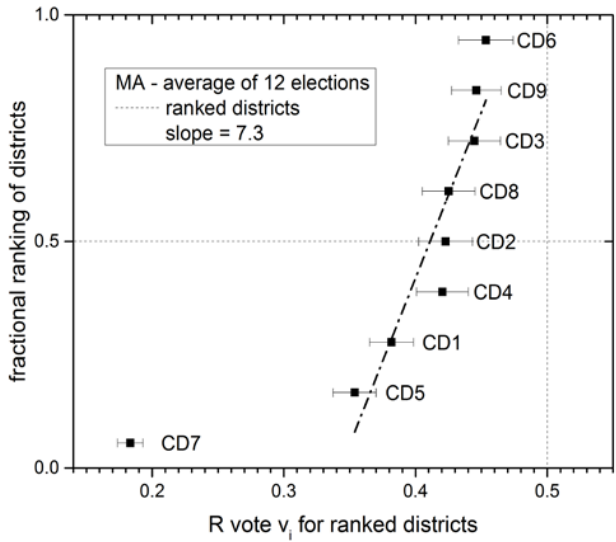
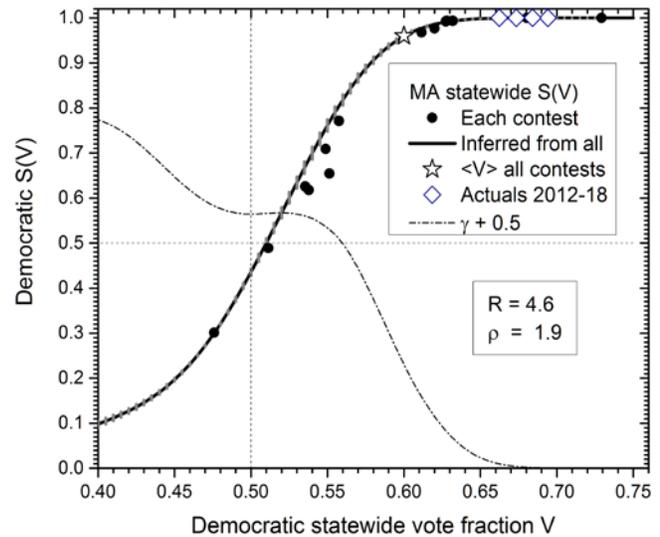

Fig. 13.  r(v) graph for Massachusetts         Fig. 14.  S(V) graph for Massachusetts

## 5. Additional ideals and measures

The MA and MD comparison just described illustrates the difficulty with assessing bias in unbalanced states. We have pursued this issue in Appendix C by considering the impact of political geography. While we are enthusiastic about the concepts described there, the difficulty in carrying out the associated procedure motivates our consideration in this section of simpler measures in addition to the $\alpha_S$, $\alpha_V$, and $\beta$ measures already introduced.  Each of these additional measures implies an ideal form for zero bias that will be characterized.

### 5. 1  Symmetry and the global GS measure

The concept that a district plan should treat parties symmetrically is appealing, so we have been disappointed that the $\beta$ measure applied to S(V) curves fails as emphasized in Section 3.2. However, there are other ways to measure symmetry besides focusing just on the average statewide vote at <V> and its counterfactual at 1 - <V> as is done in the $\beta$ measure.  In Appendix D we describe a way to look at symmetry in the r(v) graphs.  However, that does not lead to a measure that we use in this paper, so here we go directly to a measure of bias that uses global symmetry of the S(V) graph.

---
[35] McGann, et al. 2016,77 and McDonald, et al. 2018,321 also noted a Republican bias in the MA plan.

The ideal for global symmetry is simply that β(V) = 0 for all V. To see how global symmetry bias (GS) is measured for plans that do not have global symmetry, consider the example in Fig. 17. The solid line in Fig. 17 shows the S(V) curve for Democrats in South Carolina and the dashed line shows the S(V) curve for Republicans. The two curves differ considerably. Recall that we defined ΔS(V) as half the difference $S_D(V) - S_R(V)$ because then ΔS(½) is the $α_S$ value of bias and ΔS(<V>) is the β value. In Fig. 17 β is much smaller than $α_S$ because <V> is close to the point where the S(V) curves for the two parties cross.

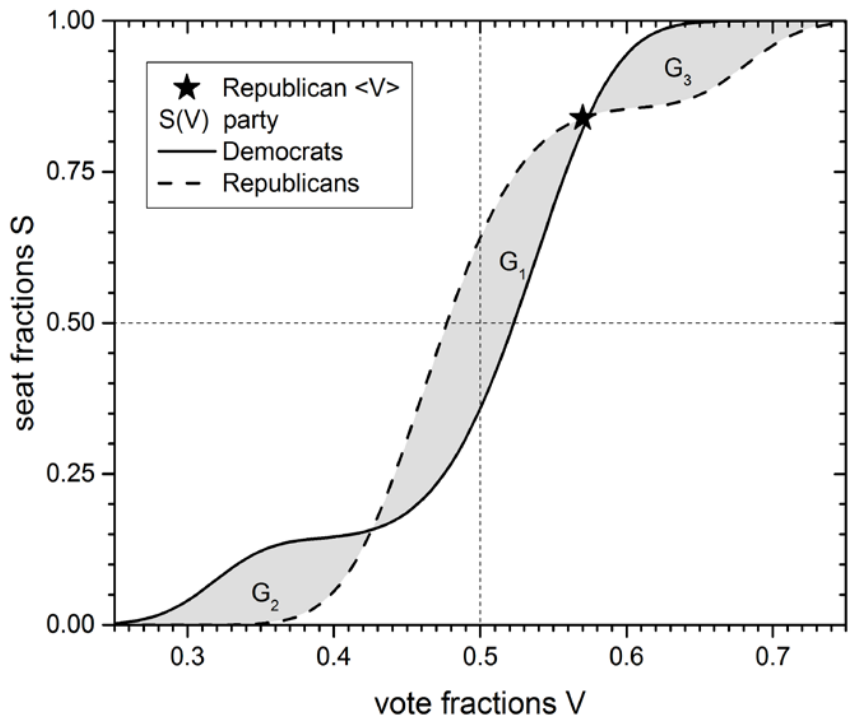

Fig. 17. Comparison of the S(V) curves for the two parties in South Carolina. The solid line shows the Democratic seat fraction versus the Democratic two-party vote fraction and the dashed line is the corresponding curve for Republicans. Each curve is the inversion of the other. Half their difference at V = 0.5 is the $α_S$ value of bias and at V = <V> it is the β value of bias. The grey area between the two curves is the GS value of bias.

If it were not for our analysis of the r(v) graphs in Section 3.2, one might say that it is unclear from Fig. 17 that the $α_S$ measure is more valid than the β measure. However, in the case of uniform shifts the area identified by $G_1$ in the figure where ΔS(V) is negative has identical size to the sum of the two areas identified by $G_2$ and $G_3$ where $\Delta S(V)$ is positive.[36] Therefore, when the $\alpha_S$ value is large, there must be crossover values of $V_c \neq \frac{1}{2}$ at which $\Delta S(V_c) = 0$. In the cases of SC, MD, and TN, <V> is close to a crossover $V_c$. Similar crossover points at similar values of $V_c$ also occur in states like NC, OH, and PA that are clearly biased; this is a further indication that $\alpha_S$ is a better measure of bias than $\beta$.[37] Nevertheless, it would be better not to have to choose a value of V at which to measure bias, especially when the chosen V is substantially different from <V>.

The total shaded area $GS = G_1 + G_2 + G_3$ between the two S(V) curves in Fig. 17 provides such an alternative measure of bias that avoids this criticism of the $\alpha_S$ and $\beta$ measures.[38] Because it integrates over all values of the vote V, it does not depend on making a choice of the value of V as the other seats based measures do. Because the difference between the two curves can also be viewed as the differences in votes needed to obtain seats, GS is best described as a combination of seats-based and votes-based measures.[39]

**5.2 New γ measure of bias**

A major concern with the $\alpha_S$ and $\alpha_V$ measures of bias for unbalanced states is that they do not evaluate near the average statewide vote <V>. The next three subsections turn to three measures of bias that do evaluate exclusively at <V> and that therefore do not employ counterfactual V.

This subsection elaborates on our γ measure newly introduced in Section 3.1 with values reported in Table 1 and shown graphically in various Figs. 6, 8, 12 and 14. It is rooted on the basic fairness principle of half the seats for half the vote V. It calculates an ideal fraction of seats

---

[36] When the proportional shift is used to construct the S(V) curves, the equivalence of these areas is not exact, but it is still approximately true.

[37] Likewise, the $\alpha_V$ votes measure of bias is better than evaluating the difference in votes required to obtain the same number of seats at the statewide <V>.

[38] GS was previously named the geometric measure of bias $B_G$ (Nagle, 2015, 351) as it is the percentage of the geometric area within the total seats/votes box defined by 0 to 1 on both axes. Prof. Grofman has kindly brought to our attention that GS is a special case of one of the eight measures in Grofman 1983, 308.

[39] Because the integral of the signed difference between the two curves is zero in the case of uniform shift and small even for proportional shift and fractional seats, GS is defined to take the absolute value of the difference which is the total shaded area in Fig. 17. Which party is favored by the necessarily positive value of GS is then taken to favor the party favored by $\alpha_S$ and $\alpha_V$. This can lead to ambiguity of the sign when the difference between the two curves is small, but such cases would be deemed acceptably fair with either sign.

$$S_\rho = \tfrac{1}{2} + \rho(<V> - \tfrac{1}{2}) \quad . \quad (5.2.1)$$

Eq. (5.2.1) is rather like the ideal fraction of seats calculated by traditional proportional representation (PR) and by the efficiency gap (EG) except that PR imposes the value $\rho = 1$ and the efficiency gap imposes $\rho = 2$. In contrast, the $\gamma$ measure uses the responsiveness determined at $V = <V>$, thereby avoiding an arbitrary choice of the $\rho$ factor, instead basing it on an empirical value at $<V>$. Then, similarly to PR and EG, the $\gamma$ measure of bias is the difference between the ideal fraction and the measured fraction

$$\gamma = S_\rho - S(<V>) = \tfrac{1}{2} + \rho(<V> - \tfrac{1}{2}) - S(<V>) \quad . \quad (5.2.2)$$

As $\gamma$ depends upon both responsiveness and seats, it could be described as a responsiveness & seats or R&S bias, although writing the simple $\gamma$ symbol is more convenient. It is also useful to recall the definition of $\mathcal{R} = (S(<V>) - \tfrac{1}{2})/(<V> - \tfrac{1}{2})$ and rewrite

$$\gamma = (\rho - \mathcal{R})(<V> - \tfrac{1}{2}) \quad . \quad (5.2.3)$$

Another favorable feature of the $\gamma$ measure is that it especially penalizes bipartisan gerrymandering, which is characterized by a smaller value of $\rho$. In unbalanced states $\mathcal{R}$ is generally larger than $\rho$. For Democratic states with $<V>$ greater than $\tfrac{1}{2}$, a smaller $\rho$ makes $\gamma$ even more negative, appropriately reporting more bias. For Republican states with $<V>$ less than $\tfrac{1}{2}$, a smaller $\rho$ makes $\gamma$ even more positive, again appropriately reporting more bias.[40]

Next, we consider what kind of S(V) curves give $\gamma = 0$. An important class is unbiased linear curves that pass through the origin with whatever slope $\mathcal{R} = \rho$. CA, IL and CO are essentially in this class. However, a general symmetric S(V) curve will have values of $\gamma$ that vary with V, usually with $\gamma$ becoming negative as $<V>$ increases from $\tfrac{1}{2}$. Of greater concern is non-symmetric curves which may have $\gamma = 0$ at special values of $<V>$.[41] The most problematic of our states in this regard is MA. Figure 14 shows that $\gamma(V) = 0$ at $V = 0.56$ and, as V increases to $<V> = 0.6$, it indicates large negative bias in favor of Democrats. Although one might be inclined to welcome this as a commonsense result, note that the S(V) curve eventually has to curve over at large V which necessarily gives a large negative $\gamma$. In the case of MA this occurs already at $<V>$, so we are inclined to discount the γ result for MA.  However, for other unbalanced, biased states like TN, γ(V) has nearly the same value for all V near <V> because S(V) has a nearly constant slope in that region.  We will return to a quantitative analysis of the γ measure for other states in Section 6.3.

---

[40] This effect also applies in balanced states.
[41] This occurs when a straight line drawn from (0.5,0.5) is tangent to the S(V) curve at V.

### 5.3  Declination measure δ

This measure (Warrington 2017, 2019) utilizes the two solid lines that we have included in our r(v) graphs. Although the difference in angles δ between these two lines does not translate to the seats and votes quantities that most people would like to know, we like this construct because it visually illustrates walls of safe seats that are characteristic of unfair maps.   In this subsection we review what the ideal declination (δ = 0) requires for S(V) partly because it has interesting similarities and contrasts with the measure in the next subsection.

The basic definition of the declination depends upon just a few quantities in the r(v) graph. Figure 18 illustrates these quantities; $r_A$ and $r_B$ are the average ranks of those districts that are won by parties A and B respectively, and $v_A$ and $v_B$ are the corresponding vote shares.[42] Equating the angles of the two lines results in the ideal δ = 0.  When the statewide vote for party A is $V_A$, straightforward algebra derives the following ideal δ = 0 relations (Campisi, et al. 2019,375, Katz, et al. 2020,175)

$$(½ - v_B) - (v_A - ½) = ½ - V_A \qquad (5.3.1)$$

and

$$S_A - ½ = (V_A - ½)/[(4v_A - 2) - (2V_A - 1)] \quad . \quad (5.3.2)$$

Equation (5.3.2) emphasizes that the seat fraction required for unbiased plans by the δ measure depends not only upon the vote $V_A$ but also upon an additional characteristic, the average vote $v_A$.[43]  Figure 19 shows the ideal S(V) curves for two values of $v_A$, 0.6 and 0.7.

---

[42] The rank of the middle circle is twice the average rank $r_B$ of seats won by party B, so this is $S_B$. Note that the district votes won by a party can have many distributions, such as (i) all having the same value or (ii) all being quite different. The lines themselves are suggestive that the district votes are linear with their rank, but this is not generally required, although such linear r(v) graphs do result in the declination not varying with the statewide vote when the uniform shift is employed. As noted earlier, we prefer to use fractional seats and vote shares in our calculations for Table 1, but it is convenient to use the simpler all-or-nothing seat assignment method (Warrington 2017)  in Figure 19 and our r(v) graphs.

[43] Or $v_B$, but that is determined from Eq. (7.3.1) once $V_A$ and $v_A$ are given.

Figure 18. Definitions of quantities $r_A$, $v_A$, $r_B$ and $v_B$ used to calculate ideal seats values for the $\delta$ measure of bias. The values shown are the same as those for NC in Fig. 3.

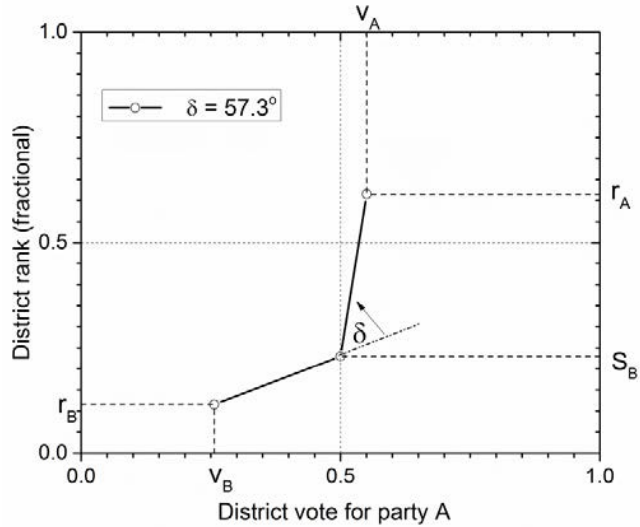

Figure 19. Party A S(V) curves for ideal declination ($\delta = 0$) and ideal lopsided outcomes (LO=0, next subsection) for two values of $v_A$ which is the average voter preference in districts won by party A.

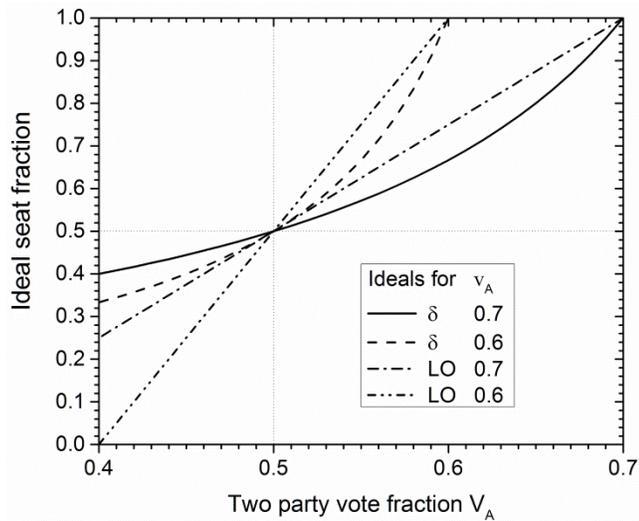

Figure 19 shows that party A wins all the districts if $v_A = 0.6$ (dashed curve) compared to 2/3 of the districts if $v_A = 0.7$ (solid curve).[44] Any measure that allows this kind of variation means that measure violates a primary principle for measuring bias (McGhee 2017, Cover 2018), namely, that a partisan map drawer not be allowed to increase the seat fraction for a given vote fraction without incurring a change in the value given by the measure.[45] In the case of the declination measure, when party A is in power and expects $V_A > ½$, it can increase $S_A$ for the same $V_A$ by decreasing $v_A$ with no change in $\delta$. Fortunately, decreasing $v_A$ also increases the competitiveness of the districts and a partisan party A map drawer would be unlikely to reduce $v_A$ low enough to create competitive districts when party A is the majority party. Unfortunately, when $V_A < ½$, Fig. 19 shows that party A obtains more seats by having all districts more packed, in which case party A would encourage a bipartisan gerrymander while satisfying $\delta = 0$.

---

[44] Using Eq. (7.3.1) the corresponding packing fractions 1- $v_B$ for party B are 0.5 and 0.4 for $v_A = 0.6$ and 0.7, respectfully. These numbers emphasize that packing of party B is not allowed by the declination ideal.

[45] Katz, et al. 2020,175 have similarly criticized the $\delta$ measure.

## 5.4 Lopsided outcomes (LO) measure of bias

This measure focuses on discriminatory packing. The ideal for this measure is that the excess vote share for districts won by party A averaged over those districts (namely, $v_A - ½$ in the previous section) equals the excess vote share for districts won by party B averaged over those districts (namely $½ - v_B$). Then, the measure of packing bias is[46]

$$LO = (½ - v_B) - (v_A - ½) \ . \qquad (5.4.1)$$

A positive value of LO would indicate greater packing of party B voters and would therefore indicate a bias in favor of party A.

The ideal LO = 0 and Eq. (5.4.1) require that $v_B = 1 - v_A$. Together with the relation $V_A = v_A S_A + v_B S_B$, simple algebra gives the seats-votes relation,

$$S_A - ½ = (V_A - ½)/2(v_A - ½) \qquad . \qquad (5.4.2)$$

Figure 19 shows the S(V) curves for two values of $v_A$. Just as for the $\delta$ measure in the previous subsection, given statewide vote $V_A$ in an unbalanced state with $V_A > ½$, a partisan map drawer could increase party A seat fraction by decreasing $v_A$, so the LO measure, like the $\delta$ measure, violates the principle that a map drawer should not be able to change the seat outcome without incurring a change in the value given by the measure of bias.[47] Although this LO measure is similar to the $\delta$ measure in the sense that both of their ideal curves depend upon $v_A$, Fig. 19 shows that they have different ideal curves.[48] The ideal S(V) curve obtained from the LO measure is similar to the proportionality and EG measures in that it is linear. Furthermore, its ideal responsiveness in Eq. (5.4.2) is $\rho = R = ½(v_A - ½)$; this relates to accommodating political geography which would be expected to produce smaller values of $v_A - \frac{1}{2}$ for more homogeneous states. However, the actual value of LO is given by Eq. (5.4.1) and that depends upon $\frac{1}{2} - v_B$ which is a separate variable for states with LO $\neq 0$.

---

[46] The usual calculation for each party simply adds the excess vote fractions just in those districts won the party and divides by the number of those districts. We use fractional seats and fractional excess votes for responsive districts, although differences are usually small.

[47] Similarly to the ideal $\delta$ measure, party A could increase $v_A$ in order to increase $S_A$ when $V_A < ½$, Katz, et al. 2020,174 have also published Eq. (5.4.2) and criticized the LO measure.

[48] This is also easily seen from Eq. (5.3.1) which is required for $\delta = 0$ whereas LO = 0 requires the left-hand side of Eq. (5.3.1) to be zero, and these two requirements differ except when $V_A = ½$.

## 5.5 Minimal inverse responsiveness $\zeta$

We turn in this subsection from fairness to responsiveness because many reformers believe that districts should be competitive. This would make it easier for voters to change the party in power, and it would provide for more robust elections. It is therefore pertinent to consider maximal responsiveness as an ideal and to devise a measure of the extent to which a map does not satisfy the ideal. Such a formula has to take into account a maximum ideal value $\rho_{max}$. Clearly the maximum occurs when all the districts j have partisan preferences $v_j = 0.5$. Of course, this is only possible when the statewide vote is $V = 0.5$. If one uses all-or-nothing assignment of a district's seat, then a small shift in statewide vote flips all the districts; that means $\rho_{max} = \infty$. Infinity is not a subtractable number and that makes it awkward to quantify how a real map differs from the most responsive. To accommodate this, we define an inverse responsiveness measure $\zeta$ as:

$$\zeta = (1/\rho) - (1/\rho_{max}) \qquad . \qquad (5.5.1)$$

Ideal responsiveness is then indicated by $\zeta = 0$, and more responsive maps are identified as having smaller values of $\zeta$. With all-or-nothing district seat assignment $1/\rho_{max}$ is simply zero. However, when the vote swings by a small amount, it is quite unlikely that all the fully competitive districts will swing to the same party, so it is preferable to calculate $\rho_{max}$ using fractional seats. For the fractional seats function we have used in this paper, $\rho_{max} = 10$. We consider this value of $\rho_{max}$ to be a reasonable maximum responsiveness for balanced states. This reduces $\zeta$ by 0.1 in Eq. (5.5.1) compared to using the assignment of just 0 or 1 to a district's seat.

The ideal $\rho_{max}$ should be modified when a state is unbalanced.[49] One might then retain as many fully competitive $v_j = \frac{1}{2}$ districts as possible and pack the remaining districts with as large a fraction of the majority party voters as possible. Figure 20 shows an example of a corresponding r(v) graph. Let us suppose that the maximum average packing of majority voters is $v_M$ and that this is large enough for those districts to be completely safe. The responsiveness is then the product of the fraction $S_C$ of competitive districts and the responsiveness of the fully competitive district with partisan preference ½. When the party A vote is $V_A$, one finds $S_C = (v_M - V_A)/(v_M - ½)$.[50] For example, for an unbalanced state that has $V_A = 0.6$ and $v_m = 0.7$, $S_C$ decreases to ½ and the $1/\rho_{max}$ term in Eq. (5.5.1) increases to 0.2.

---

[49] In this general case, it becomes more complicated when the fractional district seat is determined as a continuous function of v. For simplicity in this paragraph, we will simply use all-or-nothing assignment to determine the seats when the statewide vote is $V_A$, with $\frac{1}{2}$ assigned to fully competitive districts with partisan preference $\frac{1}{2}$.

It is also interesting that this model for maximal responsiveness gives the relation

$$S_A - ½ = (V_A - ½)/(2v_M - 1) \qquad (5.5.2)$$

when $v_M$ is large enough for completely safe districts. This gives winners bonus $R = 1/(2v_M - 1)$. In the unrealistic case of being able to pack districts with all of the majority party voters (*i.e.* $v_M = 1$), one has $R = 1$, like the proportional representation ideal. For a more realistic value of $v_M$ like 0.75, then $R = 2$, like the EG ideal.

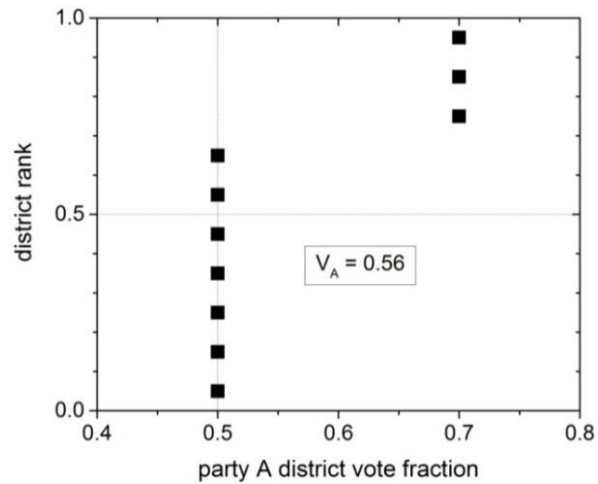

Fig. 20 Suggested rank graph to maximize responsiveness $\rho$ for a state with 10 districts, maximal average majority party packing $v_M = 0.7$, and average statewide vote $V_A = 0.56$.

## 6. Comparisons

Now that data have been shown and various measures have been described, we come to the most important part of this paper in subsection 6.1 where we synthesize our results to measure total partisan bias in unbalanced states. Comparison of measures that don't work well is given in subsection 6.2, and the durability of measures is shown in section 6.3.

---

[50] This follows directly from $V_A = ½S_C + v_M(1-S_C)$.

## 6.1 Best measures of partisan bias in unbalanced states

Different measures of bias measure different quantities. The $\alpha_S$, $\beta$ and $\gamma$ measures focus on seat bias in the S(V) graph, the $\alpha_V$ measure focuses on vote bias, and GS measures a combination of the two. Regarding the r(v) graphs, the LO measure focuses on packing districts, the declination $\delta$ measure is a geometric quantity that relates to both packing and cracking, and like the $\alpha_V$ measure, the more traditional median minus mean (mM) (McDonald and Best 2015), focuses on vote bias. Questions naturally arise: Which of these measures is best? And is there any agreement among them? These questions cannot be answered by examining a single map for a single state. In this subsection we show that they can be answered by comparing the performance of measures across multiple states.[51]

The comparison is facilitated by normalizing the values obtained by the measures to a common scale. This gives the graphical comparison shown in Fig. 21. In this figure the $\alpha_S$, $\delta$, $\gamma$ and GS values were each normalized to the $\alpha_V$ values, so the meaning of the scale in the vertical axis is the percentage of the Democratic vote in excess of 50% for half the seats.[52] The average of these normalized measures is shown by the stars. We call this composite bias and assign it the symbol $\Omega$. Values of $\Omega$ are given in Table 1.

The normalized measures shown in Fig. 21 are visually well correlated as is confirmed by the r values in the legend.[53] All the signs agree except for the MA $\gamma$ outlier which has been omitted from the statistics for the reasons given in Section 5.2. We hypothesize that the composite average $\Omega$ of these measures provides a better estimate of bias than any one of them. The t-test obtains the 99% confidence ranges for $\Omega$ shown by the uncertainty bars in Fig. 21. These indicate substantial partisan bias for Republicans in PA, NC, OH, SC, TN and TX. At the 95% confidence level (not shown in Fig. 21), MD and CA are biased in favor of Democrats and MA and IL are biased in favor of Republicans.[54]

---

[51] Although there have been many studies comparing states, we are unaware of this particular application to assess measures of partisan bias. We also note that one could also address these questions by comparing many maps for the same state.

[52] The normalization factor for each measure was obtained by regression of its values to the $\alpha_2$ values. Standardizing disparate measures to obtain an average has also been accomplished by Stephanopoulos and Warshaw 2019.

[53] The pairwise comparisons of the 5 measures are also well correlated.

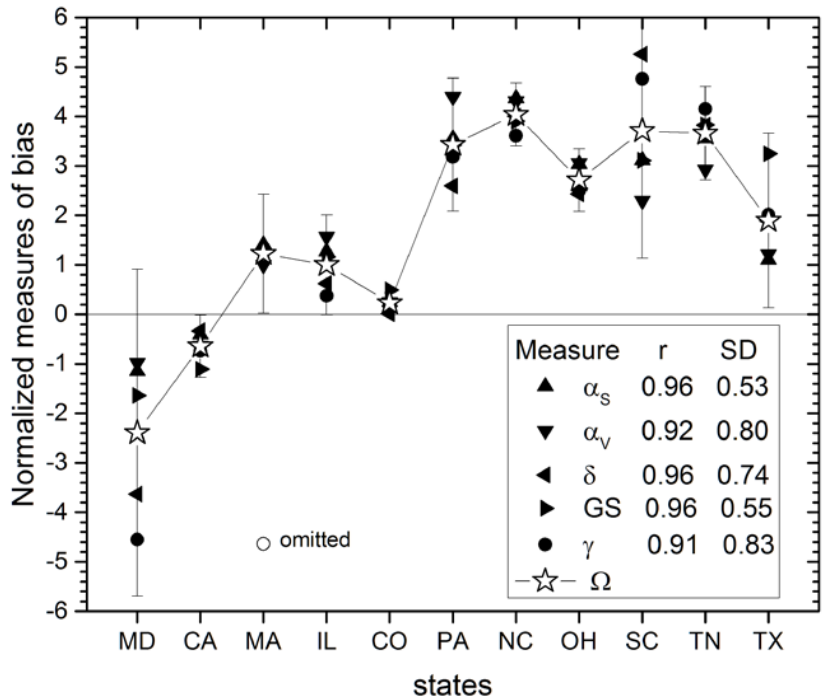

Fig. 21. Comparison of bias across states for five measures normalized to the $\alpha_V$ measure. Mean values are shown by stars with 99% confidence ranges. The legend shows the values for each measure of the Pearson r correlation and the standard deviation relative to the state mean averaged over states.

For each measure the legend in Fig. 21 also shows the average value over all states of its standard deviation (SD) relative to the stars in Fig. 21. This individual SD of each measure provides an estimate of its uncertainty when it is applied separately from the other measures. At the 95% confidence level of two standard deviations, these SD values would indicate bias for all five of these measures in PA, NC, OH, SC and TN and none of them indicates bias in CO and CA. In MD only the $\alpha_V$ measure falls short of the 95% confidence level. For the three states MA, IL and TX, some of the measures fall on either side of, but usually close to the 95% confidence level. These results suggest that any one of these measures could often be an adequate measure of bias whenever it is inconvenient to calculate all of them. When all five can be calculated, we suggest using the composite value $\Omega$.

---

[54] When comparing two plans for the same state, the uncertainties are likely to be correlated for each measure which would reduce the uncertainty for discriminating their relative bias ranges.

### 6.2 Measures that don't work in unbalanced states

Measures that disagree numerically from the averages shown in Fig. 21 are shown in Fig. 22. The β value agrees well for balanced states with the starred average because <V> is nearly the 50% that is used for the α measures. Fig. 22 also shows values of β close to zero for unbalanced states which simply reiterates the point made in Section 3.2 that there has to be a crossover point between the S(V) curves for D and R in the unbalanced range as indicated in Fig. 17.

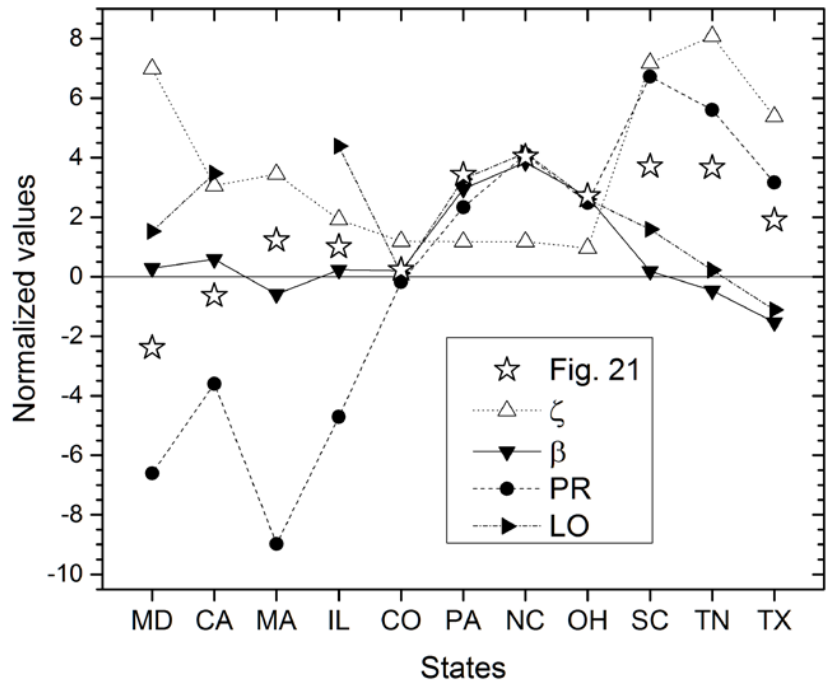

Fig. 22. Comparison across states for four measures and the average from Fig. 21, normalized to the $\alpha_V$ measure.

The PR values derived from proportionality (see Appendix C.1 for definition) in Fig. 22 also agree with the other measures for balanced states, but they differ substantially in unbalanced states which we attribute to political geography. Such deviations would be smaller if accurate estimates would be made of $PR_{SMD}$ in Eq. (C.2.1). Of course, PR is very strongly correlated with the winner's bonus R in unbalanced states.[55]

The LO values shown in Fig. 22 are consistent with those of other measures in balanced states by indicating much more bias in favor of Republicans in NC, OH and PA compared to CO. It has previously been noted (Wang, et al. 2018,313) that LO is not suited for detecting bias in unbalanced states. Indeed, the value of LO indicates a bias in favor of Republicans in MD and an even larger Republican bias in IL and CA. The Republican LO bias for CA is easily seen from its r(v) graph in Fig. 9. Because CA has essentially symmetrical r(v) and S(V) graphs, average packing is necessarily smaller in the R won districts than in the D won districts.[56] In TX, LO gives a bias in favor of Democrats. Both these unlikely values suggest that LO contains a spurious piece for unbalanced states which makes it report too much bias in favor of the minority party.[57]

---

[55] In the Democratic majority states, the appropriate correlation is between R values and negative PR values.

Values of $\zeta$ would not be expected to correlate well with any of the other values because $\zeta$ is a measure of responsiveness, not of partisan bias.[58] Surprisingly, the heavily biased balanced state maps such as PA, NC and OH have smaller, more favorable values of $\zeta$ than unbalanced states. This is due to their statewide Democratic vote occurring for $<V_D>$ slightly greater than ½ where the S(V) curve rapidly rises. Evaluating $\rho$ at smaller $<V_D>$ where the actual congressional votes occurred would make $\zeta$ larger for these states. Even so, $\zeta$ would not be correlated with measures of bias in the unbalanced states.

## 6.3 Durability

It is desirable that the numerical values given by a measure of bias do not change much for typical maps when the vote swings by plausible percentages around the best estimate for the statewide average $<V>$. Many measures of bias change dramatically with small changes in the vote when all-or-nothing assignment of a district's seat is used for states with a small number of districts. We will call this instability. It does not reflect the true character of a plan or of a measure of bias. Instability is appropriately avoided by calculating fractional seats.

This subsection discusses a more fundamental type of change which is often described as durability (Grofman 2019, Wang, et al. 2018) and as sensitivity (Warrington 2019). We define a measure to be absolutely durable if there is no change in the bias as the vote swings.[59] An example of an absolutely durable measure of bias is the $\alpha_S$ measure. Although $\alpha_S$ is evaluated at

---

[56] This follows from the definition of LO with respect to the difference in separation of the outer open circles from the inner open circle in the r(v) graphs.
[57] By comparison of Eq. (5.4.1) with Eq. (5.3.1), $<V_D>$ - ½ should be subtracted from LO.
[58] Bipartisan gerrymandering reduces responsiveness but does not affect partisan fairness in balanced states.
[59] This definition neglects systematic temporal variation differentially by precinct and district. Although the used data do not allow us to address precinct level changes, differential temporal variation of districts is contraindicated by the relative stability of their rankings over different elections.

V = ½, that evaluation is performed on the S(V) curve which doesn't change after it is drawn for a map using many past elections.[60] Likewise, the $\alpha_V$ and GS measures do not depend on choosing a statewide vote. Durability has also been a strong argument in favor of the median minus mean (mM) measure which uses the r(v) graph.[61]

Other measures of bias are not absolutely durable. Non-durability of the δ, PR, EG, ζ, LO and γ measures is examined for NC in Fig. 23 and for TN in Fig. 24. These two figures illustrate that none of these measures is absolutely durable for all states.[62] While the LO measure is most durable for NC, it is least durable for TN. The PR and γ measures are most durable for TN, but PR is least durable for NC while Fig. 12 shows durability concerns for γ in MA. The δ measure is fairly durable for TN, although it varies more for NC. Of course, one way to superficially make these measures durable would be to average over a range of V centered on <V>. However, such averaging tends to bring the value of bias back to what is obtained just using the central value obtained at <V>, so the intrinsic non-durability remains. This argues in favor of using the durable measures of bias mentioned in the previous paragraph. The ζ measure of the departure from ideal responsiveness defined by Eq. (5.5.1) is not constant in Figs. 23 and 24, simply because the S(V) curves are not linear. For CA, which has nearly linear S(V) and r(v) graphs, the γ, ζ and δ measures are quite durable. The Ω measure is the most durable measure in Figs. 23 and 24.

---

[60] The S(V) curves we calculate are quite insensitive to large variations in V as seen by the small uncertainties in our S(V) figures.
[61] The mM measure is durable insofar as the median district vote swings equally with the statewide vote when the uniform shift is employed, and that is only marginally changed when proportional shift is employed for typical swings. The mM values do depend delicately on a single median district compared to the conceptually similar $\alpha_V$ measure, so we include only the latter in our set of core measures.
[62] The numerical values of the different measures are quite different which leads us to normalize all the values to unity at V = <V>. An exception is made for LO in TN because its value is nearly zero at <V>; the graph shows 1+10LO, which emphasizes that LO changes sign in this range of V.

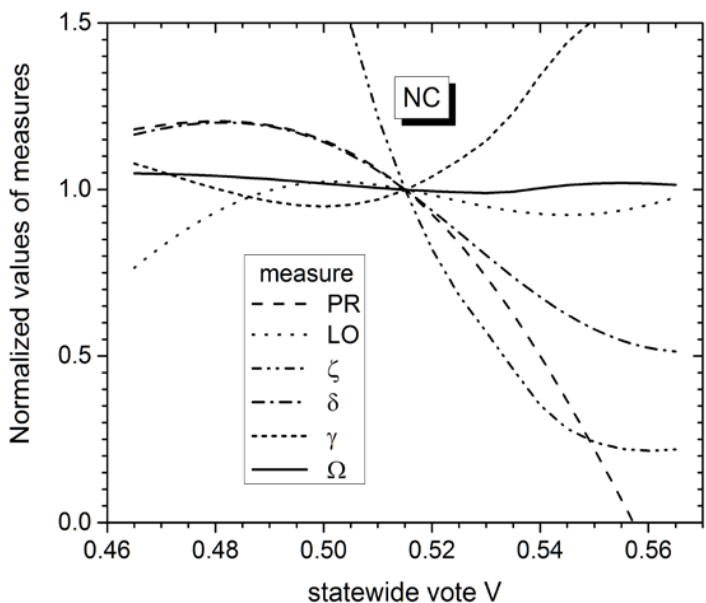

Figure 23. Non-durability of five measures of partisan bias and the $\zeta$ measure of responsiveness applied to the 2011 enacted map for NC, normalized to 1 at $V = <V>$. Actual values at $<V>$ are shown in Table 1.

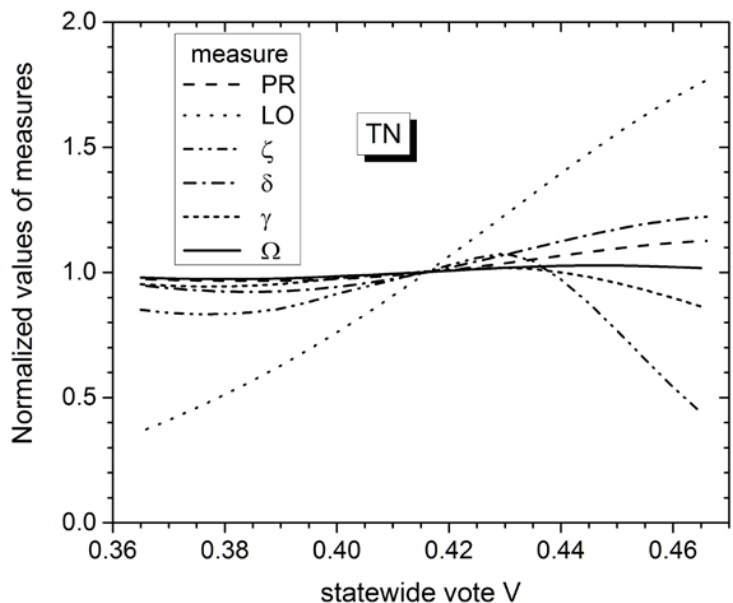

Figure 24. Non-durability of five measures of partisan bias and the $\zeta$ measure of responsiveness applied to the 2011 enacted map for TN, normalized to 1 at $V = <V>$ except for the LO measure (see FN 62). Actual values at $<V>$ are shown in Table 1.

## 7. Discussion

We preface this discussion by reminding the reader that our measures of bias are for total partisan bias, whatever its provenance. Overtly intended gerrymandering is just one part. Another part is un-remediated bias due to a state's political geography and its map-drawing rules.

### 7.1 Balanced states

Although the primary focus of this paper is on unbalanced states that lean strongly towards one party, it is important to emphasize that the thorny issues that arise for such states do not much affect balanced states. It does not matter whether one prefers proportionality (PR) or the efficiency gap (EG) (McGhee 2014) or the cubic law (Kendall and Stuart 1950) or bilogits (King 1989); when the vote share is close to 50%, they all converge on the unassailable criterion of 50% seat share for 50% vote share. While vote share is never precisely 50%, we have shown that many measures of bias agree rather well with each other for the 2011 plans of the four nearly balanced states CO, PA, NC, and OH.

Fig. 21 shows this agreement for the seats-based measure $\alpha_S$, the votes-based measure $\alpha_V$, the geometric declination measure $\delta$, a global symmetry measure GS, and a new responsiveness & seats-based measure $\gamma$. Figure 22 extends this agreement to the LO packing measure, the proportionality measure PR, and the extended symmetry measure $\beta$.[63] This agreement is revealed in these figures by normalizing the values obtained in Table 1. Different measures provide values for different quantities, such as angles for the $\delta$ measure, seats for the $\alpha_S$ measure, and votes for the $\alpha_V$ measure. These quantities provide different perspectives on the underlying bias. These quantities naturally have different scales, nearly 60 degrees for $\delta$ and only 5% for $\alpha_V$. Normalization provides a common scale that allows one more easily to compare different measures.

The agreement in this paper of many measures of bias for balanced states confirms the unsurprising conclusion that the enacted 2011 plans in NC, OH, and PA are highly biased in favor of Republicans. The same analysis also reveals that the CO plan is quite fair. All this reassures map drawers that measuring bias in balanced states is quite achievable with a wide range of measures.

### 7.2 Measures of bias for unbalanced states

The preceding fine agreement of the normalized values obtained for all measures of bias when applied to balanced states only holds for some of those measures when applied to unbalanced states. Figure 21 in Section 6.1 shows agreement for the seats-based measure $\alpha_S$, the votes-based measure $\alpha_V$, the geometric declination measure $\delta$, a global symmetry measure GS, and a new measure $\gamma$ that uses both seats and responsiveness at the average vote $<V>$. We will call these the core measures and we propose that any one of these could be used to measure bias in unbalanced states.[64] This is a key finding in this paper.

---

[63] One can add the efficiency gap measure EG to this list as its bias values are very near the same as the proportionality PR values for these balanced states.

While the $\alpha_S$, $\alpha_V$, GS, and $\gamma$ measures all use the S(V) curves that shift the primary r(v) data and employ fractional seats, the $\delta$ measure relies only on the primary r(v) data.[65] The $\delta$ measure ideal is subject to manipulation as shown by Fig. 19, but it is best at quantifying walls of safe, but not packed, districts favoring one party that are visually apparent in the r(v) graphs. As shown in Figs. 23 and 24 in Section 6.3, values of bias obtained from the $\delta$ and $\gamma$ measure depend upon the average vote $<V>$, so they are free of the vote shifting counterfactual. However, they are not as durable as the other three S(V) based measures. The $\alpha_S$ and the $\alpha_V$ values are absolutely durable but only because they are evaluated near the center of the S(V) curve, which is displaced from the statewide vote for unbalanced states. However, the S(V) curves for unbalanced states like SC, TN and TX that have strongly biased values of $\alpha_S$ and $\alpha_V$ have quite similar shape to those of NC, OH and PA whose values of $\alpha_S$ and $\alpha_V$ clearly show bias according to the fundamental principle that half the seats should be obtained for half the vote. The $\alpha_S$ and $\gamma$ measures give a relatable quantity, seat bias, and the $\alpha_V$ measure gives vote bias, whereas the $\delta$ and GS quantities are more abstract. The normative principle for the $\alpha_S$, $\alpha_V$ and $\gamma$ measures is the fundamental half the seats for half the vote, whereas the normative principal for the $\delta$ and GS measures is symmetry. GS measures absolute asymmetry over the whole range of the S(V) curve, but which party is biased against requires appealing to one of the other measures. Clearly, none of these five measures is the silver bullet for evaluating bias in unbalanced states. A natural question is: Why do these measures generally agree with each other? We suggest that each, in its own way and albeit imperfectly, is getting at the core underlying bias in a plan. We propose that together, like buckshot, they are sufficient for the job at hand, namely, to show that measuring bias is not limited to balanced states. This then leads us to consider a suitably normalized average, the composite measure $\Omega$, which we suggest is better than any of the individual measures from which it is composed.[66]

---

[64] It is, of course, logically possible that all five measures could be mutually wrong, but this appears unlikely in view of overall agreement of the results with *a priori* intuition.

[65] Nagle 2019 has criticized the $\delta$ measure for being unstable, but our use of fractional seats makes it stable. Warrington 2019 has instead proposed adding fictitious buffer districts to achieve stability.

### 7.3 Two rejected measures of bias for unbalanced states

The $\beta$ measure of bias is based on a symmetry principle that seems so plausible that it came as a surprise that it does not work in states with a dominant party. Fig. 22 shows that $\beta$ has small values for all the unbalanced states. In section 3.2 we show how and why this occurs based on examination of the $r(v)$ graphs. Those graphs are highly asymmetric as seen by their values of $\delta$, but they lead to $S(V)$ curves that are necessarily rather symmetric, but only near the value $<V>$ for unbalanced states. For other values of V, especially those for which values of $\alpha_S$, $\alpha_V$ and GS are obtained, the $S(V)$ curves are highly asymmetric. That is why these are better measures of bias than $\beta$. Withdrawing confidence in the $\beta$ measure does not repudiate the concept that symmetry between parties is still an ideal. It simply means that a broader notion of symmetry is required. Appendix D defines district symmetry (DS) in the $r(v)$ graph, and the GS measure uses global deviation from symmetry in the $S(V)$ graph. Of course, the $\delta$ measure directly illustrates deviations from symmetry in the $r(v)$ data. The $\alpha_S$ and $\alpha_V$ measures zoom in on the central portion of the $S(V)$ curve where asymmetry is robustly apparent. The $\gamma$ measure only uses the symmetry principle at the 50/50 point along with counterfactual-free data.

The LO motivation of focusing on packing looks attractive *a priori*. However, values of LO bias in Fig. 22 would claim that CA and IL are strongly biased in favor of Republicans while TN is fair and TX is biased in favor of Democrats, disagreeing strongly with our core measures and common perceptions. Section 5.4 argues that this indicates a systematic artifact that assigns too much bias in favor of the minority party in unbalanced states.

### 7.4 Proportional Ideal

In Appendix C.1 an argument is reiterated that proportionality is the ideal fair representation for partisan groups of voters and that the PR measure defined as S - V is the appropriate measure of bias. But then in Appendix C.2 the well-known incompatibility of proportionality and the single member district system due to partisan geography is reiterated. In particular, MA is an example of a state where relative geographic uniformity of partisan voters makes it essentially impossible to draw congressional districts that would durably give Republicans a proportional share.

---

[66] Averaging the five basic measures into $\Omega$ also provides estimates of uncertainty as shown in Fig. 21. Notice that these uncertainties are different than the generally smaller statistical uncertainties in the five basic measures which arise from the different statewide elections.

A redistricting procedure based on the proportionality ideal would be to find plans that come closest to satisfying proportionality for the most likely vote share $<V>$.[67] As emphasized by the example of PA (Nagle 2019), a serious drawback is a state imposing constraints on the set of acceptable plans that makes it impossible to achieve a fair map. Furthermore, the ensuing $PR_P$ estimate of bias defined in Appendix C.2 can't be obtained just from a plan itself but requires drawing many other possible maps. That is impracticable and certainly limits the number of people who could evaluate maps they have drawn.

Of course, researchers are now capable of generating ensembles of many maps by using various computer simulation techniques. As described in the introduction, often that is done to establish benchmarks or baselines that give average values of bias and then measuring intentional bias as differences from the baseline average. This has been used for court cases to evaluate intentional gerrymandering, but that is not what $PR_p$ in Section C.2 attempts to measure. It measures bias as the difference from the plan that comes closest to proportionality, and that putatively fairest plan may even be an extreme outlier in the set of computer generated plans when a state's political geography favors one party.[68] This is consistent with the quote at the end of section 4.1 from McDonald, et al. 2018,323 to the effect that the Illinois legislature drew an outlier map in 2011 that turned out to be rather fair. The challenge with implementing this approach is in determining what is the fairest possible plan with respect to proportionality. For that reason, in the body of this paper we have retreated to our five core measures $\alpha_S$. $\alpha_V$. $\delta$. $\gamma$ and GS and their composite $\Omega$ that appear to measure total bias reliably without this complication and the related complication of having to consider political geography and its interaction with the state's rules for map drawing.

---

[67] Since there would be other maps that would grant the majority party a larger winner's bonus R, we are aware that it may be deemed naïve even to suggest such a criterion. However, a criterion that would guarantee minority party seats, by packing minority party voters if necessary, has recently been favorably discussed by Katz, et al. 2020, Appendix B.

[68] Cf. the introduction and footnote 3. Also, note that this section applies equally to other measures like the efficiency gap that has an ideal S(V) curve.

### 7.5 Responsiveness

We have considered several quantities related to responsiveness. The basic one is ρ; it measures how responsive the plan is at the expected value of the statewide preference <V>. Bipartisan gerrymanders that lock in safe seats for either party have small values of ρ, so it is an important measure for reformers. It is most easily measured from the S(V) curve. Table 1 shows values of ρ that are suspiciously small for MD, SC, TN, and TX. We have also defined a reciprocal responsiveness measure ζ in section 5.5 that allows comparison of a plan to a realistic maximum responsiveness that takes into account the difference between unbalanced and balanced states. This responsiveness measure is not well correlated with bias, reaffirming that fairness and responsiveness are two separate quantities (King and Browning 1987).

An overall responsiveness R is the relevant quantity for assessing anti-majoritarian outcome in balanced states and the winner's bonus in unbalanced states. In the latter case a large value of R is associated with unfairness to the minority party as in MA, MD, SC, and TN, but it likely has a minimum value that depends on the political geography of a state. This measure of responsiveness is therefore entangled with bias in both balanced and unbalanced states.

The ratio ρ/R is also of interest; a ratio near 1 in unbalanced states, such as CA and IL, follows from their linear district symmetry (DS) as seen in Appendix D. Bipartisan gerrymandering in unbalanced states reduces ρ/R by reducing ρ. This ratio is also small in balanced states that are biased because R is generally large and even negative for anti-majoritarian states like PA, NC, and OH.

Another measure of responsiveness that we have not focused on is the fraction of competitive districts. A rough way to estimate this is to multiply ρ by a competitive range of votes, typically 10%.[69] It is somewhat discouraging that this means that there would only be 20% competitive districts even with ρ = 2 super proportional responsiveness. In any case, the S(V) curve is the most appropriate vehicle to assess responsiveness/competitiveness.

### 7.6 Future application to redistricting

---

[69] Better than a fixed range with a sharp cutoff would be a gradual fractional district measure. While competitiveness at the statewide level and responsiveness have distinct measures, they are clearly closely related.

In order to focus on the thorny issue of measuring bias in unbalanced states, this paper has performed an analysis across states using the single 2011 plan for each state. Moving forward to redistricting in 2021, we propose using five core measures in all states for the purpose of comparing the bias in different proposed plans. Although any one of these measures would provide an estimate of the total bias, a better estimate would average their normalized values as in our composite $\Omega$ measure. The normalization factors we obtain from our analysis across states are, for normalization to the $\alpha_V$ measure, 0.21 for $\alpha_S$, 0.083 for $\delta$, 0.66 for GS, and 0.17 for $\gamma$. These would provide a first approximation for new maps in a state. However, it is quite likely that these normalization factors will be different for different states. For example, 12 maps were analyzed for $\alpha_S$ and $\alpha_V$ using the same 7s election data in Table 1 in Nagle 2019,69. Regression on these data gives 0.31 for the normalization of $\alpha_S$ to $\alpha_V$, rather larger than the above value of 0.21. Therefore, as more maps are drawn, more precise values of the normalization factors can be obtained.[70] The main result of this paper remains the identification of the core measures of bias for all states and the analysis that devises a composite measure $\Omega$.

## 8. Conclusion

Based on the results in this paper, we propose that total partisan bias can be measured reliably for the problematic unbalanced states as well as for balanced states. Any or all of the five core measures or their composite $\Omega$ measure can be used with past election results to assess partisan bias in maps drawn by redistricting commissions, in maps drawn by individuals using various map drawing software, and in maps drawn by computer algorithms. Furthermore, responsiveness can also be obtained immediately from S(V) graphs. Measures of fairness and responsiveness that are perceived to be reliable would encourage the enactment of election law that would include these fundamental concepts, and in a form that could have justiciable bite.

---

[70] Furthermore, following up on footnote 54, estimates in the difference of the bias between two plans should be addressed using Student's t-test on the differences in the values of the five normalized measures.

# Appendix A. Data

## A.1 Source.

The data come from statewide election returns compiled and disaggregated to voter tabulation districts (VTDs) by Stephen Wolf at Daily Kos (Wolf 2014). Since U.S. elections at all levels are administered by county or local governments, Wolf disaggregated county level returns to assign votes to VTDs. While this is not ideal, first order discrepancies were avoided by using available votes cast in the VTD in the 2008 presidential election and the proportion of the county's population living in a VTD. McDonald, et al. 2018 have reported that Wolf's disaggregation method was sufficiently accurate by comparison to VTD level data that were available in several states. Even if Wolf's data were deemed unsuitable to determine bias for the actual states in question, these data are sufficient to test measures of bias as this can be done for any consistent set of data.

| State | 2004 President | 2006 Senate | 2006 Governor | 2006 Attorney General | 2006 Secretary of State | 2006 Comptroller | 2006 Treasurer | 2006 Auditor | 2008 President | 2008 Senate | 2008 Attorney General | 2008 Secretary of State | 2008 Treasurer | 2010 Senate | 2010 Governor | 2010 Attorney General | 2010 Secretary of State | 2010 Comptroller | 2010 Treasurer | 2010 Auditor | 2012 President | 2012 Senate | 2012 Governor | 2012 Attorney General | 2012 Secretary of State | 2012 Treasurer | 2012 Auditor | 2013 Senate (special) |
|---|---|---|---|---|---|---|---|---|---|---|---|---|---|---|---|---|---|---|---|---|---|---|---|---|---|---|---|---|
| CA | | | | | | | | | ✓ | | | | | | ✓ | ✓ | ✓ | ✓ | ✓ | | ✓ | ✓ | | | | | | |
| CO | ✓ | | ✓ | ✓ | ✓ | | | | ✓ | ✓ | | | | ✓ | ✓ | ✓ | ✓ | | | | ✓ | | | | | | | |
| IL | ✓ | | ✓ | ✓ | ✓ | ✓ | ✓ | | ✓ | ✓ | | | | ✓ | ✓ | ✓ | ✓ | ✓ | ✓ | | ✓ | | | | | | | |
| MA | ✓ | ✓ | ✓ | ✓ | | | | | ✓ | ✓ | | | | ✓ | ✓ | ✓ | ✓ | | ✓ | ✓ | ✓ | ✓ | | | | | | ✓ |
| MD | ✓ | ✓ | ✓ | ✓ | | ✓ | | | ✓ | | | | | ✓ | ✓ | | ✓ | | | | ✓ | ✓ | | | | | | |
| NC | ✓ | | | | | | | | ✓ | ✓ | | ✓ | ✓ | ✓ | ✓ | ✓ | | | | ✓ | ✓ | | ✓ | | ✓ | ✓ | ✓ | |
| OH | ✓ | ✓ | ✓ | ✓ | ✓ | | ✓ | ✓ | ✓ | | ✓ | | | ✓ | ✓ | ✓ | ✓ | | ✓ | ✓ | ✓ | ✓ | | | | | | |
| PA | ✓ | ✓ | ✓ | ✓ | | | | | ✓ | | | | | ✓ | ✓ | | | | | | ✓ | ✓ | | ✓ | | | | |
| SC | ✓ | | ✓ | | ✓ | ✓ | ✓ | | ✓ | ✓ | | | | ✓ | ✓ | ✓ | ✓ | ✓ | | | ✓ | | | | | | | |
| TN | ✓ | ✓ | ✓ | | | | | | ✓ | ✓ | | | | ✓ | | | | | | | ✓ | ✓ | | | | | | |
| TX | ✓ | ✓ | ✓ | ✓ | | ✓ | | | ✓ | ✓ | | | | ✓ | ✓ | | | | | | ✓ | ✓ | | | | | | |

Table A.1 Used statewide elections for each state as well as a downballot composite for each state.

The states that were used are given in Table A.1 Our initial choice of states to analyze was to follow McDonald, et al. 2018 who chose the four states, IL, MA, MD, all unbalanced D states, along with balanced OH. We then wanted to have a comparable number of unbalanced R states, choosing TX as the largest one to match CA which we thought would be valuable as the widely touted most fairly districted state, as well as two smaller unbalanced R states, TN and SC, to match MA and MD. We also added notoriously biased balanced states PA and NC as well as CO which looked *a priori* to be the least biased balanced state. The used elections and results are in the research/elections directory of this GitHub repository: https://github.com/dra2020/nagle. Elections that had significant third-party votes were not used.

### A.2. Calculations for r(v) and S(V) graphs

For each congressional district j in each election we calculated the two-party Democratic vote fraction $v_j$ = D vote / (D vote plus R vote). For each district, the average vote fraction and the standard error of the mean was calculated and plotted in the r(v) graphs. The statewide two-party vote fraction in each election was calculated in two ways: 1) the average of the district vote fractions and 2) the actual two-party vote fractions in the state. The difference of 1) minus 2) is the turnout bias (McDonald 2009). Table A.2 shows values for the CA elections that we used. Table A.3 shows average values of turnout bias for all analyzed states.

| Election | 08P | Down | 10AG | 10C | 10G | 10SS | 10T | 12P | 12S | Mean | STD |
|---|---|---|---|---|---|---|---|---|---|---|---|
| <V> % | 62.1 | 59.7 | 50.4 | 60.4 | 56.7 | 58.2 | 60.9 | 61.9 | 62.5 | 59.2 | 3.8 |
| t-bias % | -0.2 | -0.6 | -0.7 | -0.8 | -0.8 | -0.9 | -0.8 | -0.6 | -0.5 | -0.7 | 0.2 |

Table A.2 Average statewide vote <V> and turnout bias for CA elections

| state | CA | CO | IL | MA | MD | NC | OH | PA | SC | TN | TX |
|---|---|---|---|---|---|---|---|---|---|---|---|
| % t-bias | 0.7 | -0.3 | 0.6 | 0.7 | 0.4 | 0.2 | 0.3 | -0.2 | 0.1 | 0.3 | 2.2 |
| % std | 0.2 | 0.1 | 0.3 | 0.4 | 0.2 | 0.3 | 0.2 | 0.2 | 0.2 | 0.2 | 0.3 |

Table A.3 Average turnout bias for all states.

Because turnout bias is small, it has been ignored in the main text where the tables and figures show the actual two-party vote fractions.

For each election $\varepsilon$ a seats/vote $S_\varepsilon(V)$ graph was calculated for statewide vote $V_k$ from 25% to 75% in intervals of 0.5%. For each $V_k$ a voter preference $v_j(V_k)$ was calculated for each district j using a proportional shift from the factual $v_j(<V>)$ as defined by Nagle 2015, 2019. When the shifted statewide party preference $V_k$ for party A is smaller than <V>, party A voters in each district are shifted to party B. The proportional shift shifts the same fraction of party A voters in each district. In contrast, the uniform shift shifts the same fraction of total voters from party A to party B regardless of how few party A voters there may be.[71] The fraction of a seat estimated for each shifted district was then calculated using party seat probability $P(V) = 1 - \frac{1}{2}(1 + \text{prob}((V - \frac{1}{2})/0.04))$ where prob is the usual probit function, here with variance 0.04.[72] The sum of seat fractions over all districts was then $S_\varepsilon(V)$ for that election. Figure A.1 shows $S_\varepsilon(V)$ curves for several elections in CA. Then, the S(V) curve is the average of the $S_\varepsilon(V)$ curves and the error bars in the S(V) graphs are the standard error of the means at each V. Finally, solid circles in the S(V) figures show $S_\varepsilon(V_\varepsilon)$ where, for each statewide election $\varepsilon$, $V_\varepsilon$ is the statewide vote and $S_\varepsilon$ is the average sum of district fractional seats. Differences between the individual $S_\varepsilon(V_\varepsilon)$ and $S(V_\varepsilon)$ are another measure of the uncertainty of our estimates.

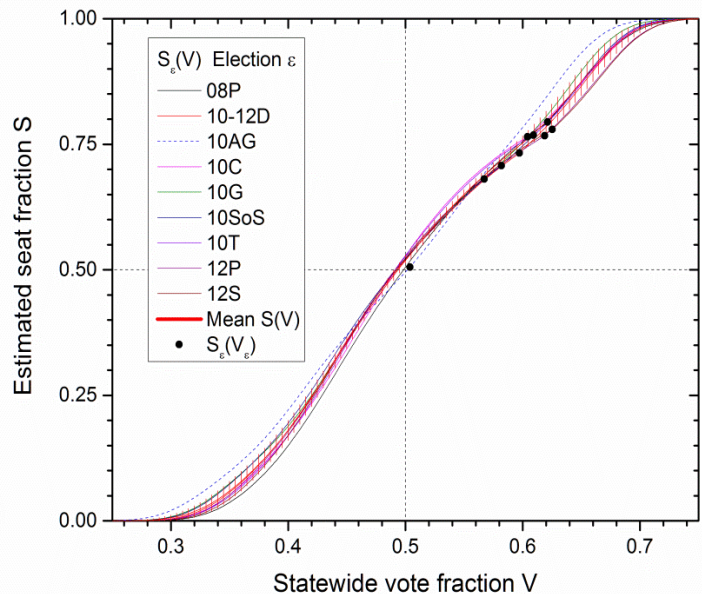

Fig. A.1 Solid circles show seats $S_\varepsilon(V_\varepsilon)$ for nine CA elections $\varepsilon$ from whose district votes are obtained the shifted Democratic $S_\varepsilon(V)$ curves. Averaging gives the mean S(V) curve shown with standard deviations.

## Appendix B. Comparison of proportional and uniform methods for shifting vote shares

It has long been recognized that reliable extrapolations can be made by shifting the vote that is obtained in actual elections. The simplest way to do this, much used over the years, is the uniform shift method which shifts every precinct and every district by the same percentage as the shift V - <V> in the statewide vote. While this time-honored method is likely to be reliable for competitive districts j with $v_j$ close to 0.5, it has the obvious flaw for unbalanced districts that $v_j$ may become greater than 1 or less than 0 (King 1989), so we employ the proportional shift method. This is based on the more plausible model that it is equally likely that any voter anywhere in the state is equally likely to shift his/her vote in the same proportion as the statewide shift. Of course, this is a model. It is quite likely that some precincts have such strong partisan preferences that hardly anyone will vote differently when the statewide vote changes and other precincts will be more responsive than average. We have done a cursory analysis that suggests that such differences tend to average out at the level of congressional districts when many elections are considered, but that is beyond the scope of this paper. The purpose of this appendix is simply to show how the S(V) curves compare for the two different models.[73]

---

[71] The uniform shift can obviously result in districts with less than 0% or greater than 100% preferences, but that can't happen with the proportional shift.
[72] For example, this function estimates that a district with 55% preference for party A has a seat likelihood of 10.5% for party B as shown in Fig. 1 in Nagle 2019.

Figure B.1 shows the comparison for PA. As expected, both methods agree with each other and with the actual election results in the actual vote range. The difference for large shifts is also expected. The uniformly shifted S(V) only goes to 0 when the statewide shift away from Democrats is large enough to shift the most Democratic districts to become safe Republican (this also shifts some Republican districts to have more than 100% Republican voters). In contrast, as the statewide vote shifts against Democrats using a proportional shift, the most Democratic districts lose Democratic voters more rapidly, so the Democratic proportional shift S(V) curve lies below the uniformly shifted S(V) curve for small V. Similarly, as the vote shifts toward Democrats it takes a larger statewide shift for the most Republican districts to become Democratic, so the uniform shift S(V) lies below the proportional shift S(V) for large V.

Fig. B.2 for CA shows the same differences between the uniform shift and the proportional shift as Fig. B.1. Fig. B.2 also shows $\beta(V)$ where 0.5 has been added to $\beta$ in order to centralize the curves on the same figure. The proportionally shifted $\beta(V)$ reverses sign as V increases from 0.5 where it is equal to $\alpha_S$ and slightly favors Democrats to $<V>=0.6$ where it would indicate bias in favor of Republicans, contrary to all our core measures in Section 7 as shown in Table 1 in the main text. The uniformly shifted $\beta(V)$ follows the same course with V but its magnitude is much smaller at $<V>$ than the proportional shift because S(V) is larger at the counterfactual V = 1 - $<V>$. For other states the sign is even different as shown in Table B.1. Such large

---

[73] Katz, et al. 2020,172 have recently noted that the proportional swing has three times the error of uniform swing in their data base. The errors are smaller and more nearly equal for the two methods in our analysis. We also note that their paper used alternative (1) in Sec. A.2 for the vote fraction.

differences in the results using two plausible models for shifting votes is further reason not to trust the β measure of bias. The $α_S$ measure is clearly better as Table B.1 shows that its values have the same sign in each state for both shifting methods, although the magnitudes do differ, most in TX where votes in the actual elections require the greatest extrapolation to V = 50%.

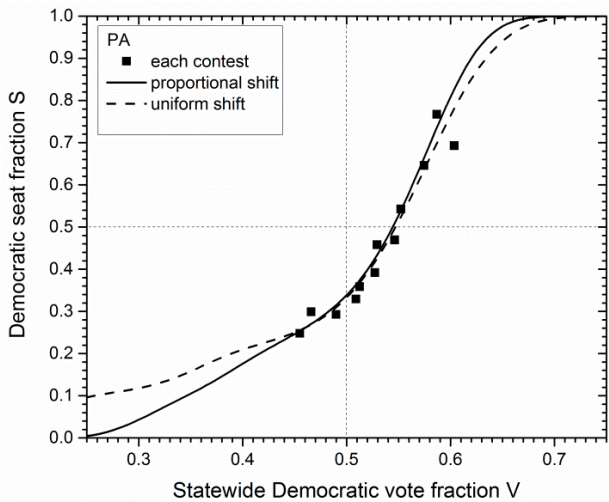 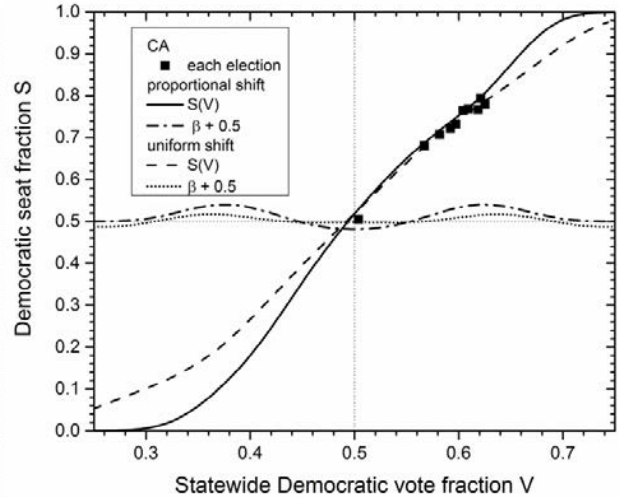

Fig. B.1  S(V) for two shift models in PA　　　　Fig. B.2.  S(V) for two shift models in CA

| state | CA | IL | MA | MD | SC | TN | TX |
|---|---|---|---|---|---|---|---|
| beta values | -- | -- | -- | -- | -- | -- | -- |
| prop shift | 2.9 | 1.2 | -2.9 | 1.4 | 0.9 | -2.3 | -7.5 |
| uniform shift | 0.6 | 1.6 | -3.0 | -2.1 | 4.9 | 2.2 | 1.1 |
| alpha values | -- | -- | -- | -- | -- | -- | -- |
| prop shift | -1.9 | 5.8 | 6.4 | -5.2 | 14.1 | 16.2 | 5.0 |
| uniform shift | -0.2 | 9.3 | 12.7 | -3.7 | 16.9 | 21.2 | 12.7 |

Table B.1.  Percentage values for α and β for unbalanced states as determined using proportional shift and uniform shift.

# Appendix C.  Proportionality vs. Political Geography and the SMD System

## C.1  The proportionality ideal

Proportionality is widely recognized as the ideal in countries that use a list or mixed member proportional system to determine the fraction of seats awarded to different parties, but courts have not recognized it in the United States.  Indeed, SCOTUS opinions have specifically denied that proportionality has constitutional support. A more cogent argument against proportionality is that it is not generally achievable over a wide range of the vote in the SMD system. A counterargument is that proportionality only has to be achievable over the expected range of the vote in each state. We will come back to achievability in the next subsection.

Let us we review why proportionality can be considered the ideal, even in an SMD system. We begin with the assertion that each voter should be represented equally as any other voter. Two voters of like-mind in the same district with the same representative clearly have equal representation, but that is different from the representation of a voter of opposite mind in the same district. That is an unfortunate feature of the SMD system of representation. Nevertheless, groups of voters of like-mind can still be represented equally on average in the SMD system and that is how one can argue for proportionality as the ideal.[74] In this view representatives are shared among voters, so the empowerment of each voter of like-mind is the number of representatives of the like-minded party divided by the number of voters who voted for that party. Let us designate S and V as the fraction of seats and voters for party A and 1-S and 1-V as the fractions for the other party B. Then, the share of representation per voter of the two parties is just S/V and (1-S)/(1-V) respectively. Requiring equal shares for voters in both groups and trivial algebra then requires S = V, which is just proportionality.[75] In terms of measures of responsiveness, this is R = $\rho$ = 1.

If proportionality is the democratic ideal, that suggests a simple measure of bias, namely,

$$\text{PR} = S(<V>) - <V> \ . \qquad (C.1.1)$$

This evaluation is only performed at the average vote share <V> in an attempt to avoid the achievability problem of demanding proportionality over a wide V range. The PR values for the states we have analyzed are shown in Table 1[76].

The PR measure is related to overall responsiveness R by

$$PR = (R-1)(<V> - ½), \qquad (C.1.2)$$

which shows that PR depends both on the winner's bonus R and the state imbalance $<V> - ½$.[77]

---

[74] Brief Amici Curiae of Common Cause and the Campaign Legal Center, Inc, *Shapiro v. McManus,*14-990 (U.S. Aug. 13, 2015).

[75] Note that one can also generalize this in two ways. First, even representatives of the opposite party provide representation for a voter of the opposite party, just not as much; although the algebra is more complicated, proportionality again ensues. Second, one can derive proportionality for any number of parties.

[76] Regarding the sign of PR, when Republican values are inserted for S and <V> as is done in Table 1, then positive values of PR mean bias in favor of Republicans.

Another theoretical ideal is provided by the efficiency gap (EG) principle that the wasted votes should be equal for both parties. This results in $R = \rho = 2$. Other variations have also been discussed (Nagle 2017). Historical averages give values of R near 2. This has suggested that idealizing an R =2 might take into account the effect of the SMD system. Then one has the well-known EG measure of bias (McGhee 2014),[78]

$$EG = (S(<V>) - ½) - 2(<V> - ½) . \qquad (C.1.3)$$

Like the PR measure, EG also depends upon R and $<V> - ½$ according to $EG = (R-2)(<V> - ½)$.

## C.2 Political geography and a modified proportionality measure

The end of Section 4.2 essentially asks a question: Why should MD be generally acknowledged to have been gerrymandered whereas MA is not so considered when the estimated 4% fraction of Republican seats in MA is so much smaller than the 14.3% fraction in MD for the nearly identical 60% partisan imbalance in both states? This question can be reframed in terms of the quantitative measures in the previous subsection. Why should the MD plan not be considered fairer than the MA plan when MD has smaller PR bias[79] than MA if the value PR = 0 is to be considered the ideal based on proportionality? One answer could be simply that PR may not be a good measure of bias. We next consider a more nuanced answer to this question that takes into account the contribution of political geography in the SMD system to the PR measure.

If the political complexion of an unbalanced state is homogeneous, then any map must lead to the dominant party winning all the seats. We believe that this is close to being the case for MA because we have been unable to draw even one reasonable looking Republican leaning district in MA.[80] In contrast, the 2011 MD plan clearly showed that it was possible to draw a map with a strongly Republican district. While the MD plan is fairer to its minority party voters by conforming better to proportionality than MA, one clearly can't blame the makers of the MA plan for an unfair map if it was absurd to draw even one Republican district. In the case of MA, we suggest that its large value of PR bias can mostly be attributed to its relatively homogeneous political geography interacting with the SMD system.

---

[77] Eq. (C.1.2) emphasizes how PR varies with R for a state's relative vote share. However, the factor ($<V> - ½$) is crucial when considering balanced states as its small value drives R into negative, anti-majoritarian values when the state is biased as seen for NC, OH, and PA in Table 1.
[78] Although the EG was originally expressed in terms of wasted votes, this form is equivalent when turnout bias is zero, and this is the form that properly accommodates turnout bias Cover 2018, McGhee 2017, Katz, et al. 2020.
[79] MD also has smaller $\gamma_2$ bias.

In other unbalanced states it is certainly possible to draw some districts that are favorable to the minority party, but it still may not be possible to achieve PR = 0 due to their political geography. In that case, the fairest achievable plan would be the one that minimizes the absolute value of PR given the political geography of the state. We suggest that the value of PR that is closest to zero be called $PR_{SMD}$ because it is a bias due, not to intentional gerrymandering, but to the incompatibility of the SMD system with proportionality. For maps that do not achieve the $PR_{SMD}$ value, that value could then be subtracted to obtain an effective plan bias

$$PR_P = PR - PR_{SMD}. \qquad (C.2.1)$$

$PR_P$ can then be compared for plans within each state and also to compare plans enacted by different states.

Let us illustrate how this proposal works for MA. For MA, Table 1 shows PR = − 36% which is implausibly even larger than the Republican PR advantage in balanced, clearly biased states NC, PA and OH. Let us consider as a rough estimate made from drawing a few maps and informed by the results of Duchin, et al. 2019 that about one durable and acceptable looking Republican majority district can be drawn in MA. If so, then the maximum fraction of Republican seats at $V_R$ = 40% is 1/9 and $PR_{SMD}$ = (1/9) – 0.40 = -29%. Then, the modified bias is $PR_P$ = -36% - (-29%) = − 7%. This is considerably smaller in magnitude and more realistic looking than the unmodified PR = − 36%.

This MA example indicates that the absolute values of PR for other unbalanced states in Table C.1 may also be too large and require similar modification to take account of political geography. Consistency then suggests that a similar modification should also apply to balanced states, so we turn next to PA where we uncover cautionary concerns. Substantial numbers of maps for PA were drawn for the recent court case which overturned the 2011 Congressional map. Those maps were drawn under the tight constraint that the number of county splits was to be minimized.[81] Under those constraints, Nagle 2019 estimated that the maximum number of Democratic seats would be 7.5 for 50% Democratic two-party vote .[82] One might then suggest that $PR_{SMD} = (18-7.5)/18 - 0.5 = 8\%$ using 50% vote. The problem with accepting this in Eq. (C.2.1) is that any PA plan with 58% R seats for 50% vote would be assigned zero bias ($PR_p$ = PR - $PR_{SMD}$ = 0) while strongly violating the most fundamental fairness criterion of 50% seats for 50% vote. The root cause for not being able to draw a fair PA map with more than 7.5 seats for 50% vote is simply the tight county split constraint imposed by the PA Supreme Court on acceptable maps. It was shown that loosening the county split criterion allowed a fair map to be drawn (Nagle 2019). That having been done, then the minimum absolute value of $PR_{SMD}$ is 0 and only PA plans that satisfy the fundamental 50/50 fairness criterion have $PR_p$ = 0. While this shows that consistency can be established in PA in the context of this $PR_p$ modified measure, it emphasizes the importance of the constraints imposed on acceptable maps and concern over how that choice is made.

---

[80] Duchin, et al. 2019 have shown rigorously that it is impossible to draw any district that would have elected a Republican based on 5 statewide elections in the 2002-2010 cycle. They also reported that MA has become more heterogeneous, and we have used the districtr software program to draw a contiguous, although not 'reasonable looking', district that had substantial Republican preference in all of the 8 statewide election results after 2008.

Recapitulating what we learn from the MA and PA examples, the underlying goal of the proportionality ideal is to assign bias to a plan whenever proportionality is not achieved without regard for the source of the bias. The $PR_p$ modification attempts to remove that part of the bias that can be attributed to the SMD system which usually brings about a winner's bonus R greater than the proportionality ideal R = 1; Eq. (C.1.2) shows this effect. The primary goal of the $PR_p$ modification is to enable proportionality values for unbalanced states to appear consistent with values obtained for balanced states which do not need such a modification.[83]

---

[81] PA practice requires population deviation not to exceed *one person*; then the minimal number of county splits has to be precisely the number of districts minus one, barring some highly improbable set of county populations, http://lipid.phys.cmu.edu/nagle/Technical/Theorem-splits.docx. Note also that the number of county splits is not the same as the number of split counties, *e.g.* one county split between three districts counts as two splits.

[82] The estimate for the 2018 adopted PA map was 7 D seats for 50% of the vote ($\alpha$ = 11%) and 9 seats for 53.6% of the vote ($\alpha_2$ = 3.6%) which indeed was the 2018 outcome. Cervas and Grofman 2020,9 have estimated 7.8 D seats for 50% of the vote for the adopted map.

**C.3  Survey of other states**

We have not attempted to estimate $\gamma_{SMD}$ for other states.[84] Instead, let us just qualitatively discuss which of those states might have had fairer maps. For this we will refer to a quantitative measure of heterogeneity in the political geography (PG) of states recently devised by Chen and Rodden 2018. Their primary goal is to measure the difference in PG between parties. For most states they find that the average Democratic voter is more likely than the average Republican voter to reside in a congressional sized neighborhood (Eubank and Rodden 2019) that has a high fraction of same party voters. This difference in the political geography of the parties leads to natural packing of Democratic voters and partisan bias if not remediated.[85] Chen and Rodden found that MD has the largest difference between parties, strongly favoring Republicans, and MA has one of the smallest differences. This is not exactly what we would like to know in order to understand why MA must have nearly all Democratic seats while MD need not. Nevertheless, we will use the sum of their heterogeneities for both parties as a qualitative proxy for the overall heterogeneity of a state. We will call this the PGS measure. From our discussion of MA, it is also relevant to consider just the heterogeneity of Republicans in Democratic dominated states, which we will call PGR, and the heterogeneity of Democrats in Republican dominated states, which we designate PGD. We also show values for these measures of PG in Table C.1.

The larger value of PGS for MD than for MA generally suggests that the political geography of MD is more heterogeneous than that of MA. The small value of PGS and PGR for MA is consistent with the difficulty we had drawing a single Republican district in MA. However, the PGR heterogeneity of Republicans in MD isn't very much larger than in MA, although it is far smaller than the average heterogeneity of Democrats.[86]  Although MD has one Republican district, it is packed to over 60% R, and the state's large PGS suggests that that district could be made more responsive along with one or more of the currently Democratic districts.[87] These considerations suggest that MD is more biased than MA. However, the relatively small PGR suggests that it is unlikely that the proportional ideal of 3 R seats could be achieved.[88]

---

[83] For a balanced state, Eq. (C.1.2) shows that any plausible value of the winner's bonus does not contribute significantly to PR because the <V> - ½ factor is small. For balanced states NC, OH and PA, the actual PR bias is due to the large negative anti-majoritarian values of R. We also note that a redefinition of R to be the slope of the vector from the V=50% point on the S(V) curve instead of from the (50,50) center leads one back to the $\alpha_S$ measure.

[84] Estimates of $PR_{SMD}$ will undoubtedly be obtained by minority party map drawers in the next round of redistricting.

[85] Unintentional gerrymandering can be considered to be a consequence of politically geography and the SMD system under the legal and conventional constraints imposed on drawing maps.

| state | <V> | S(<V>) | R | ρ | α1 | PR | PGD | PGR | PGS |
|---|---|---|---|---|---|---|---|---|---|
| TX | 40.4 | 27.8 | 2.3 | 1.1 | 5.0 | 12.6 | 6.9 | 1.8 | 8.7 |
| TN | 41.6 | 19.2 | 3.6 | 0.8 | 16.2 | 22.4 | 5.9 | 0.6 | 6.6 |
| SC | 43.0 | 16.1 | 4.9 | 0.9 | 14.1 | 26.9 | 1.9 | 0.1 | 1.9 |
| OH | 51.3 | 41.4 | -6.0 | 4.5 | 13.8 | 9.9 | 4.7 | -0.6 | 4.0 |
| NC | 51.5 | 35.1 | -10.0 | 4.0 | 19.8 | 16.4 | 3.3 | -0.6 | 2.7 |
| PA | 52.9 | 43.6 | -2.2 | 4.1 | 16.1 | 9.3 | 5.1 | 1.9 | 7.1 |
| CO | 50.6 | 51.3 | 2.2 | 3.9 | 1.1 | -0.7 | 4.2 | 0.3 | 4.4 |
| MD | 59.3 | 85.7 | 3.7 | 1.0 | -5.2 | -26.4 | 9.1 | 1.0 | 10.1 |
| IL | 60.0 | 78.8 | 2.9 | 3.1 | 5.8 | -18.8 | 4.2 | 4.6 | 8.8 |
| MA | 60.0 | 96.0 | 4.6 | 1.9 | 6.4 | -35.9 | 2.2 | 0.5 | 2.7 |
| CA | 59.2 | 73.6 | 2.6 | 2.1 | -1.9 | -14.4 | 3.8 | 3.2 | 7.0 |

Table C.1. The last three columns give values for the political geography of states from Chen and Rodden 2018. The PR column shows values obtained from the proportionality measure. The definitions of the preceding columns are given in the Table 1 caption. All values are percentages except for R and ρ which are the slopes shown on the S(V) graphs.

The other two unbalanced D states in Table C.1, CA and IL, both have much larger PGR and PGS than MA, consistent with their smaller values of PR. It is difficult to judge whether the political geography of CA would allow a smaller PR value. But it seems likely that the fraction of Democratic seats in IL could have been smaller, thereby reducing its PR, consistent with its larger PGR and PGS compared to CA.

Turning next to unbalanced states with substantial Republican majorities, SC has a PR value favoring Republicans and with a smaller magnitude to that of the negative PR in MD favoring Democrats. However, because SC has a larger PGD (1.9) than the PGR in MD (1.0), its $PR_{SMD}$ would be smaller so the magnitude of its $PR_P$ could be larger and more biased than MD. TN has a bit smaller PR value than SC, but its PGD value is much larger suggesting that TN is even more biased than SC. TX has the largest PGS and PGD values and the smallest PR value, so it is difficult to compare it to TN and SC. The PR value for TX is close to the magnitude of PR for CA, but the PGD is twice as large as the PGR for CA, so its $PR_P$ would be larger, suggesting more R bias in TX than D bias in CA.

---

[86] Quite likely, the problem is that averages obscure the possibility that Republicans may be packed in certain regions of the state, where a Republican district has been drawn, and diluted in other regions, leading to a relatively homogeneous average.

[87] This is basically what reformers in MD advocate.

[88] This might require even more severe packing of the two highly urban MD districts that have over 70% Democratic preference.

For the balanced states in our survey, it is noteworthy that all four have much larger values of PGD than of PGR, so PG favors Republicans as emphasized by Chen and Rodden 2018. Nevertheless, the PR value for CO is quite fair in contrast to the values for the other three balanced states. However, the latter values are quite interesting when compared to the PR values for the unbalanced states. It is particularly noteworthy that the magnitude of PR in CA is larger than in PA. Superficially, this might suggest that CA is more gerrymandered than PA. However, this does not take into account the possibility that CA requires a substantial value of $PR_{SMD}$ whereas $PR_{SMD}$ close to 0 is achievable in PA. That could make the $PR_P$ bias for CA smaller than for PA. This reversal of the comparative bias of the CA and PA plans would thus be accounted for by subtracting the bias embedded in the SMD system.

While $PR_P$ is well defined in Eq. (C.2.1), it can't be evaluated just from a plan itself because $PR_{SMD}$ is not simply obtainable except by drawing many other possible plans. This is a drawback to this way of evaluating partisan bias, although it could be overcome by generating many plans by computer. A more serious drawback noted in footnote 3 is arbitrariness in imposing constraints on which maps are acceptable to establish the value of $PR_{SMD}$.

The complications engendered by the considerations in this appendix account for our preference for the core of simple measures discussed in the main text.

### D. Symmetry in the r(v) graphs

Let us here consider a very basic definition of symmetry at the district level. When there is a district that has a stronger preference for party A than the average statewide preference, then we define district-level symmetry to require that there be another district that has the same stronger preference for party B. Let us label two such districts m and n and designate the preferences of these districts by their expected vote share for party A as $v_m$ and $v_n$. Then, if the statewide vote share is $<V>$, this pair of districts is defined to be symmetrical when $v_m - <V> = <V> - v_n$.

Symmetry for the entire state is then achieved if every district is so paired.[89] We will call this district symmetry, abbreviated DS. For our r(v) graphs, <V> is the Republican vote <$V_R$>.

The nearly linear r(v) graph for CA in Fig. 9 comes quite close to exhibiting DS, so it is not an unattainable ideal. Indeed, any r(v) graph whose districts fall on a straight line automatically exhibits DS. However, linearity is only a sufficient condition for DS, but it is not necessary. District symmetry only requires what is called inversion symmetry. Inversion about the point (r=0.5,v=<V>) transforms each point located at r and v on the r(v) graph into a point at r' = 1 - r and v' = 2<V> - v.[90] If a graph is transformed by inversion and remains identical to what it was before inversion, then we say it has DS. Fig. 15 shows an r(v) graph that has three examples that have DS. In addition to the linear one, there are two non-linear examples that are relevant for considering responsiveness. All are centered at r = 0.5 and the statewide average v = <V> which we chose to be 0.6 for the examples in Fig. 15. However, the slope is quite different for these three examples at this central point; this slope is named $\rho_{sym}$ in the figure legend. The actual responsiveness $\rho$ is essentially the slope at v = 0.5 because that is the midrange where districts most likely change party for small statewide vote swings. The legend of Fig. 15 shows values of $\rho_{sym}$ and $\rho$, as well as the winner's bonus R.[91]

Figure 16 shows the corresponding S(V) curves which, for simplicity, are obtained by uniformly shifting the curves in Fig. 15 by 0.1 to the left. Regarding responsiveness, it is interesting to compare $\rho$ for the three examples in Fig. 16. When <$V_B$> = 0.3, the plan represented by the dot-dash line has the largest slope and the most seats for minority party B. But when <$V_B$> = 0.5, the most responsive plan is the one represented by the dashed line in Fig. 16 and by open circles in Fig. 15.[92]

---

[89] For a state with an odd number of districts, at least one district would have $v_j$ = <V>.

[90] The reason for choosing the inversion center at <$V_R$> is that this becomes the center of the S(V) graph.

[91] The slope at v = 0.5 is exactly the responsiveness if one uses the uniform swing and all-or-nothing district seat assignment. Likewise, the value of the winner's bonus R is the slope of the straight line from the intersections of the curves with the v = 0.5 vertical line to the central point at r = 0.5.

[92] One can also obtain a rough measure of the number of competitive districts when <$V_A$> = 0.6 by counting the number of districts that lie within the window 0.45 to 0.55. Then, the third plan shown by triangular symbols is least responsive, although it gives the minority party the most seats. However, when we consider <V> = 0.7, that same plan is the most responsive because all the curves shift to the right by 0.1 which is the same as counting the number of responsive districts in a window which shifts to the left by 0.1 in the r(v) graph. One may also define the competitiveness of individual districts using fractional seats probabilities, such as was done by Nagle 2019.

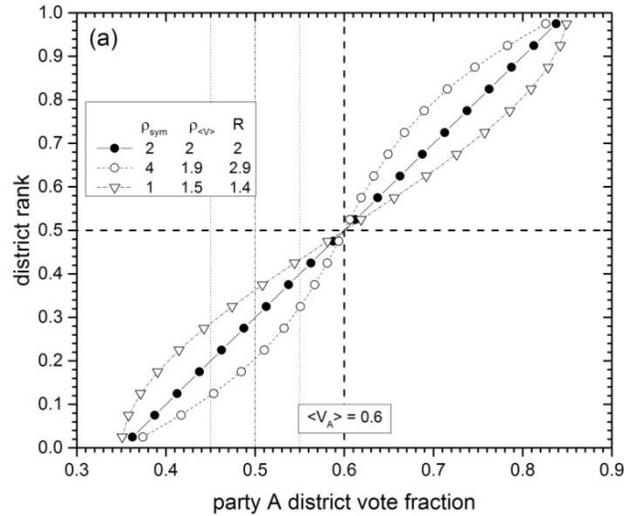

Fig. 15. Three examples of r(v) graphs that have district symmetry (DS) with the same average statewide vote $<V_A>$ = 0.6 for party A.

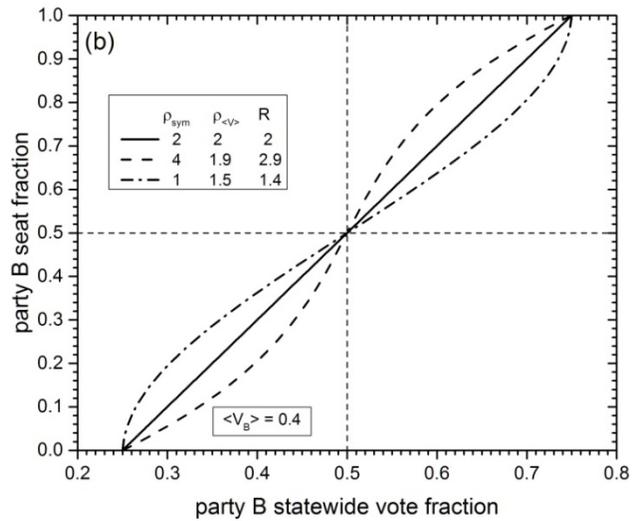

Fig. 16. The corresponding $S(V_B)$ curves assuming uniform swing of the r(v) graphs and all-or-nothing seat assignment.

**Acknowledgements:** We thank B. Grofman, J. Cervas, M. D. McDonald, G. King, G. R. J. Wallin, S.-H. Wang, M. Duchin, N. Stephanopoulos, E. McGhee, and G. S. Warrington for informative discussion. We especially thank Stephen Wolf, whose data made this study possible.

## Bibliography

Abramowitz, A. I., B. Alexander, and M. Gunning. 2006. "Incumbency, Redistricting, and the Decline of Competition in U.S. House Elections." *The Journal of Politics* 68: 75-88.
Altman, M., and M. D. McDonald. 2018. "Redistricting by Formula: An Ohio Reform Experiment." *American Politics Research* 46: 103-31.
Backstrom, C, L Robins, and S Eller. 1990. "Establishing a Statewide Electoral Effects Baseline." In *Political Gerrymandering and the Courts*, ed. B. Grofman. New York, NY: Agathon Press. 145-70.

Best, R., J. S. Krasno, D. B. Magleby, and M. D. McDonald. 2020. "Detecting Florida's Gerrymander: A Lesson in Putting First Things First." *Social Science Quarterly* 101: 37-52.

Bradlee, Dave. "Dave's Redistricting." https://davesredistricting.org.

Campisi, M., A. Padilla, T. Ratliff, and E. Veomett. 2019. "Declination as a Metric to Detect Partisan Gerrymandering." *Election Law Journal* 18: 371-87.

Cervas, J. R., and B. Grofman. 2020. "Tools for Identifying Partisan Gerrymandering with an Application to Congressional Districting in Pennsylvania." *Political Geography* 76: 102069(1-12).

Chen, J., and J. Rodden. 2018. "The Loser's Bonus: Political Geography and Minority Party Representation." *Preprint*.

Chen, J. W., and J. Rodden. 2013. "Unintentional Gerrymandering: Political Geography and Electoral Bias in Legislatures." *Quarterly Journal of Political Science* 8: 239-69.

Cottrell, D. 2019. "Using Computer Simulations to Measure the Effect of Gerrymandering on Electoral Competition in the Us Congress." *Legislative Studies Quarterly* 44: 487-514.

Cover, B. P. 2018. "Quantifying Partisan Gerrymandering: An Evaluation of the Efficiency Gap Proposal." *Stanford Law Review* 70: 1131-233.

Duchin, M., T. Gladkova, E. Henninger-Voss, B. Klingensmith, H. Newman, and H. Wheelen. 2019. "Locating the Representational Baseline: Republicans in Massachusetts." *Election Law Journal* 18: 388-401.

Eubank, N., and J. Rodden. 2019. "Who Is My Neighbor? The Spatial Efficiency of Partisanship." *Preprint*.

Gelman, A., and G. King. 1994. "A Unified Method of Evaluating Electoral Systems and Redistricting Plans." *American Journal of Political Science* 38: 514-54.

Gelman, A., G. King, and A. C. Thomas. 2012. "Judgeit Ii: A Program for Evaluating Electoral Systems and Redistricting Plans." *http://gking.harvard.edu/judgeit*.

Grofman, B. 2019. "Tests for Unconstitutional Partisan Gerrymandering in a Post-Gill World." *Election Law Journal* 18: 93-115.

Grofman, B., and G. King. 2007. "The Future of Partisan Symmetry as a Judicial Test for Partisan Gerrymandering after Lulac V. Perry." *Election Law Journal* 6: 2-35.

Grofman, Bernard. 1983. "Measures of Bias and Proportionality in Seats-Votes Relationships." *Political Methodology* 9: 295-327.

Gronke, P., and J. M. Wilson. 1999. "Competing Redistricting Plans as Evidence of Political Motives - the North Carolina Case." *American Politics Quarterly* 27: 147-76.

Katz, J. N., G. King, and E. Rosenblatt. 2020. "Theoretical Foundations and Empirical Evaluations of Partisan Fairness in District-Based Democracies." *American Political Science Review* 114: 164-78.

Kendall, M. G., and A. Stuart. 1950. "The Law of Cubic Proportion in Election Results." *British Journal of Sociology* 1: 183-97.

King, G. 1989. "Representation through Legislative Redistricting - a Stochastic-Model." *American Journal of Political Science* 33: 787-824.

King, G., and R. X. Browning. 1987. "Democratic Representation and Partisan Bias in Congressional Elections." *American Political Science Review* 81: 1251-73.

McDonald, M. D., D. B. Magleby, J. Krasno, S. J. Donahue, and R. Best. 2018. "Making a Case for Two Paths Forward in Light of Gill V. Whitford." *Election Law Journal* 17: 315-27.

McDonald, M. P. 2014. "Presidential Vote within State Legislative Districts." *State Politics & Policy Quarterly* 14: 196-204.

McDonald, Michael D. 2009. "The Arithmetic of Electoral Bias, with Applications to U.S. House Elections." *APSA 2009 Toronto Meeting Paper. Available at SSRN*.

McDonald, Michael D., and Robin E. Best. 2015. "Unfair Partisan Gerrymanders in Politics and Law: A Diagnostic Applied to Six Cases." *Election Law Journal* 14: 312-30.

McGann, A. J., A. S. Smith, M. Latner, and J. A. Keena. 2016. *Gerrymandering in America: The House of Representatives, the Supreme Court, and the Future of Popular Sovereignty*. 1st ed. New York, NY: Cambridge University Press

McGhee, E. 2017. "Measuring Efficiency in Redistricting." *Election Law Journal* 417: 426-31.

———. 2014. "Measuring Partisan Bias in Single-Member District Electoral Systems." *Legislative Studies Quarterly* 39: 55-85.

MetricGeometryandGerrymanderingGroup. "Districtr." https://districtr.org/.

Nagle, J. F. 2017. "How Competitive Should a Fair Single Member Districting Plan Be?". *Election Law Journal* 16: 196-209.

———. 2015. "Measures of Partisan Bias for Legislating Fair Elections." *Election Law Journal* 14: 346-60.

———. 2019. "What Criteria Should Be Used for Redistricting Reform?". *Election Law Journal* 18: 63-77.

Powell, R. J., J. T. Clark, and Dube. M. P. 2020. "Partisan Gerrymandering, Clustering, or Both? A New Approach to a Persistent Question." *Election Law Journal* 19: 79-100.

Rodden, J. A. 2019. *Why Cities Lose*. New York, N.Y.: Hachette Book Group.

Stephanopoulos, N. O. 2013. "The Consequences of Consequentialist Criteria." *UC Irvine Law Review* 3: 669-715.

Stephanopoulos, N. O., and E. M. McGhee. 2018. "The Measure of a Metric: The Debate over Quantifying Partisan Gerrymandering." *Stanford Law Review* 70: 1503-68.

Stephanopoulos, N. O., and C. Warshaw. 2019. "The Impact of Partisan Gerrymandering on Political Parties." http://dx.doi.org/10.2139/ssrn.3330695

Thornburgh, D. "Draw the Lines." https://drawthelinespa.org/.

Wang, S. S.-H. 2016. "Three Tests for Practical Evaluation of Partisan Gerrymandering." *Stan. L Rev.* 68: 1263-321.

Wang, S. S.-H., B. A. Remlinger, and B. Williams. 2018. "An Antidote for Gobbledygook:Organizing the Judge's Partisan Gerrymandering Toolkit into Tests of Opportunity and Outcome.". *Election Law Journal* 17: 302-14.

Warrington, G. S. 2019. "A Comparison of Partisan-Gerrymandering Measures." *Election Law Journal* 18: 262-81.

———. 2017. "Quantifying Gerrymandering Using the Vote Distribution." *Election Law Journal* 17: 39-57.

Wolf, Stephen. 2014. "Dra Update: 2012 President & Downballot Election Results Estimates + Display Table Template ". <http://www.dailykos.com/stories/2014/8/4/1318876/-DRA-Update-2012-President-Downballot-Election-Results-Estimates-Display-Table-Template>: Daily Kos.